\documentclass[twocolumn,prl,aps,psfig,showpacs,preprintnumbersm,superscriptaddress,longbibliography]{revtex4-1}
\usepackage{graphicx}
\usepackage{epstopdf}
\usepackage{float}
\usepackage{color}
\usepackage{amsmath}
\begin{document}
\title{Edwards-Anderson parameter and local Ising-nematicity in FeSe revealed via NMR spectral broadening}
\author{Paul~Wiecki}
\affiliation{Institute for Quantum Materials and Technologies, Karlsruhe Institute of Technology (KIT), 76131 Karlsruhe, Germany}
\author{Rui~Zhou}
\affiliation{Laboratoire National des Champs Magnétiques Intenses – European Magnetic Field Laboratory, UPR3228 Centre National de la Recherche Scientifique, Univ. Grenoble Alpes, Institut National des Sciences Appliquées de Toulouse, Univ. Paul Sabatier, Grenoble, France}
\author{Marc-Henri~Julien}
\affiliation{Laboratoire National des Champs Magnétiques Intenses – European Magnetic Field Laboratory, UPR3228 Centre National de la Recherche Scientifique, Univ. Grenoble Alpes, Institut National des Sciences Appliquées de Toulouse, Univ. Paul Sabatier, Grenoble, France}
\author{Anna~E.~B\"{o}hmer}
\affiliation{Institute for Quantum Materials and Technologies, Karlsruhe Institute of Technology (KIT), 76131 Karlsruhe, Germany}
\affiliation{Institut f\"ur Experimentalphysik IV, Ruhr-Universit\"at Bochum, 44801 Bochum, Germany}
\author{J{\"o}rg~Schmalian}
\affiliation{Institute for Quantum Materials and Technologies, Karlsruhe Institute of Technology (KIT), 76131 Karlsruhe, Germany}
\affiliation{Institute for Theory of Condensed Matter, Karlsruhe Institute of Technology (KIT), 76131 Karlsruhe, Germany}
\date{\today}

\begin{abstract}
The NMR spectrum of FeSe shows a dramatic broadening on cooling towards the bulk nematic phase at $T_s=90$ K, due to the formation
of a quasi-static, short-range-ordered nematic domain structure. 
However, a quantitative understanding of the NMR broadening and its relationship to the nematic susceptibility is still lacking.
Here, we show that the temperature and pressure dependence of the broadening is in 
quantitative agreement with the mean-field Edwards-Anderson parameter of an Ising-nematic model in the presence of random-field disorder introduced
by non-magnetic impurities. Furthermore, these results reconcile the interpretation of NMR and Raman spectroscopy data in FeSe
under pressure. 
\end{abstract}

\maketitle

\section{Introduction}
The nucleation of local-symmetry-breaking order in nominally symmetry-preserving phases is an increasingly recognized phenomenon in correlated electron systems. Important examples include
the formation of antiferromagnetic droplets in CeCoIn$_5$, high-$T_c$ cuprates and low dimensional quantum magnets \cite{Urbano2007,Seo_2013,Julien_2000,Kimura2003} as well as the nucleation of charge density wave order above the phase transition, in NbSe$_2$ \cite{Berthier_1978,Arguello2014,Chatterjee_2015}, ZrTe$_3$ \cite{Yue_2020,liu_2021}, cuprates \cite{Wu_2015} and  Sn/Ge(111)-$\alpha$ surfaces \cite{Weitering2107}.

In the iron-based superconductors, signatures of $C_4$ symmetry breaking and short-range nematic order have often been found well above the bulk nematic phase transition temperature \cite{Kasahara2012,Yi6878,Wang2018,Hong2020}, especially from local probe measurements using scanning tunneling microscopy (STM) \cite{Rosenthal_2014} and nuclear magnetic resonance (NMR) \cite{Iye2015,Zhou2016,Toyoda2018,Toyoda_2018b,Dioguardi2015,Dioguardi2016,Bossoni2013,Bossoni2016}.
More recently, x-ray and neutron pair distribution function (PDF) studies have revealed the locally orthorhombic nature of the 
tetragonal paramagnetic phase in the (Sr,Na)Fe$_2$As$_2$ system \cite{Frandsen2018}. Similarly, the re-entrant tetragonal antiferromagnetic state in this hole-doped system was also revealed to have short-range orthorhombic correlations \cite{Frandsen2017}. Furthermore, inelastic x-ray scattering has recently produced insights into the spatial correlation length of nematic fluctuations in iron-based 
superconductors from the wavevector-dependent softening of acoustic phonons \cite{Wu2021,Weber2018,Merritt2020,Kauth2020}. 

While the existence of local nematicity in the tetragonal phase has been rationalized in terms of residual strains in the crystal \cite{Baek2016}, it is most naturally explained 
as a consequence of impurities and vacancies that locally break the fourfold rotation symmetry and thus act as random-field impurities for nematicity.
Recent theoretical work has addressed the impact of disorder on the emergence of nematicity in iron-based superconductors \cite{Nie2014,Steffensen2019}.

FeSe is a unique iron-based superconductor ($T_c=8.5$ K) which has a bulk nematic phase below $T_s=90$ K, but no corresponding magnetic phase. 
Several experiments have revealed a large nematic susceptibility in the high-temperature phase 
\cite{Watson2015,Bohmer2015,Luo2017,Massat2016,B_hmer_2017,Hosoi_2016,bartlett2021relationship}.
Recent NMR
measurements \cite{Wiecki2017,Wang2017,Li2020} revealed a prominent broadening of the NMR spectrum on approaching $T_s$ from above, which is attributed 
to the formation of locally-nucleated, short-range-ordered nematic domains. The local orthorhombicity of unstrained FeSe above $T_s$ was later confirmed by x-ray and neutron PDF studies \cite{Koch2019,Frandsen2019} and has been used to rationalize properties of the 
tetragonal phase \cite{Wang2020}. 

Under applied hydrostatic pressure $p$, the temperature-dependent full-width-at-half-maximum (FWHM) of the 
NMR spectrum follows a universal curve, interrupted only at the nematic transition temperature $T_s(p)$. 
This pressure-independent behavior of the NMR FWHM has been interpreted as evidence that
the nematic fluctuations are robust against pressure application, despite the decrease of $T_s(p)$ \cite{Wiecki2017,Wang2017}.
However, this conclusion is in conflict with Raman spectroscopy measurements \cite{Massat2018} which
show a rapid suppression of nematic fluctuations with increasing pressure. 

In order to resolve this discrepancy it is important to establish a quantitative understanding of the 
nematic broadening of the NMR spectrum and, in particular, its relation to the nematic susceptibility. 
Here, we find that the broadening of the NMR spectrum due to locally-nucleated nematic order is proportional
to the Edwards-Anderson parameter of a random-field Ising model at the mean-field level. 
Within this picture, the pressure-independence of the NMR FWHM is seen to be a   consequence of the 
pressure-independence of random-field defects. We conclude that the NMR data are consistent with a suppression
of nematic coupling with increasing pressure and are not in conflict with the Raman data.

\section{Theory}
We consider an Ising nematic system characterized by the pseudo-spin variable $\tau_i^z$ on a lattice with site $i$. Our analysis is independent of the microscopic origin of nematicity and equally applies to systems with spin-induced nematicity \cite{Xu2008,Fang2008,Fernandes2010,Fernandes2014}, nematicity due to orbital ordering \cite{Krueger2009,Lv2009,Chen2010,Kontani2010}, or systems with a Pomeranchuk instability in the $\ell=2$ angular momentum channel \cite{Oganesyan2001,Yamase2005,Chubukov2016}. In a clean system a finite expectation value $\phi\equiv \langle \tau^z \rangle$ corresponds to nematic order. 

We allow for random strain fields $\sim h_i$ that locally break the four-fold symmetry.  To be specific we consider  random strain  characterized by the distribution function 
\begin{equation}
p_\sigma(h_i)=\frac{1}{\sqrt{2\pi}\sigma}e^{-\frac{h_i^2}{2\sigma^2}}.
\end{equation}
The width $\sigma$ is an energy scale that parameterizes the disorder strength. 
The order parameter is now $\overline{\phi}=\overline{\langle\tau^z\rangle}$, where $\langle\rangle$ is a thermal average and the overbar
is an average over disorder configurations. Despite the random strain one still expects a sharp nematic transition above which $\overline{\phi}=0$, at least for three-dimensional systems and not too strong disorder~\cite{Nattermann1988,Toh1992}.
However, for the problem at hand one expects at any finite randomness a finite Edwards-Anderson  parameter 
 \begin{equation}
 q_{EA}=\overline{\phi^2}-\left(\overline{\phi}\right)^2,
 \label{EA}
 \end{equation}
that characterizes the strength of local disorder variations, even if $\phi$ vanishes on the average.
Notice that this is different from the behavior in spin glasses where $q_{EA}$ serves, at least within mean field theory, as ``order" parameter of the spin glass state \cite{spinglass}. Nevertheless, there are interesting analogies between our approach and the theory of NMR in spin glasses or relaxor ferroelectrics, where the determination of the Edwards-Anderson parameter has played an important role \cite{Blinc1990,Chen1991,Papantopoulos1994,Blinc1999}.
In what follows we will first discuss that $q_{EA}$ is directly related to the width of the NMR spectrum. In a second step we present and solve a simple mean-field model that allows for remarkable agreement between theory and experiment.

\subsection{Connection to NMR spectrum broadening}
We start our analysis with a  brief  discussion  of the effect of random strain on the NMR spectrum. In a homogeneous case, the NMR spectrum of a system where the nucleus under consideration has one unique position in the lattice and occurs only in one isotope configuration
can be expressed as 
\begin{equation}
f(\omega)=\frac{1}{N}\sum_{i=1}^N\delta(\omega-\omega_0),
\end{equation}
where $\omega_0$ is the Larmor frequency and $N$ is the number of nuclei in the sample. 
We assume that a defect-nucleated local order parameter $\phi_i(h_i)$ will give rise to a shift of the local NMR resonance frequency according to $\omega_0\to\omega_0+\alpha\phi_i(h_i)$.
By averaging each site over the disorder distribution $p_\sigma(h_i)$ one obtains the NMR spectrum as
\begin{equation}
\overline{f(\omega)}=\frac{1}{N}\int\prod_{j=1}^Ndh_jp_\sigma(h_j)\sum_{i=1}^N\delta(\omega-\omega_0-\alpha\phi_i(h_i)).
\end{equation}
The broadening $\nu$ of the NMR spectrum is given by the second moment of the distribution $\nu^2=\int d\omega\overline{f(\omega)}(\omega-\omega_0)^2$.
Carrying through the integrations yields 
\begin{align}
\nu^2&=\frac{1}{N}\int\prod_{j=1}^Ndh_jp_\sigma(h_j)\sum_{i=1}^N\alpha^2\phi_i^2(h_i)\\
&=\alpha^2\overline{\phi^2}=\alpha^2q_{EA}.
\end{align}
Therefore, the FWHM of the NMR spectrum is expected to be proportional to the square root of the Edwards-Anderson parameter:
\begin{equation}
\nu=\alpha q_{EA}^{1/2}.
\label{eqn:FWHMqEA}
\end{equation}
This analysis was performed under the assumption of vanishing averaged order parameter $\overline{\phi}$. At finite $\overline{\phi}$ one easily finds that $\nu^2$ is proportional to the expression for $q_{EA}$ in Eq.~\eqref{EA}.

\subsection{Random-Field Ising-nematic model}
In order to get a quantitative understanding of the temperature and strain dependence of the Edwards-Anderson parameter $q_{EA}$ of Eq.~\eqref{EA}, we now perform a simple mean-field analysis of the corresponding random field Ising model. While the random-field Ising model is a theoretical problem of formidable complexity, we will confine ourselves to a  mean-field analysis. As we will see, this already allows for a rather detailed understanding of the temperature and pressure dependence of the NMR broadening for FeSe.
Mean field behavior of the nematic degrees of freedom is expected for clean systems. As was discussed in Ref.~\cite{Karahasanovic2016}, long range strain forces drive the statistical mechanics of the system into  a mean field regime in the entire temperature window where 
an appreciable softening of the shear modulus is observed.
In systems with random strain, the situation is more subtle~\cite{Nattermann1988,Toh1992} and one expects disorder fluctuations at long distances and time scales. However on the local length scale of the  measurement  of the NMR line-width, a mean field analysis is a reasonable starting point. The description of dynamical phenomena, as determined by the NMR relaxation rate, is likely more subtle and may requiring to go beyond the mean field theory.

The Hamiltonian is expressed as
\begin{equation}
H=-\sum_{i,j}J_{i,j} \tau_i^z\tau_j^z-\sum_i(h_0+h_i)\tau_i^z,
\end{equation}
where $h_0$ is an external strain that breaks $C_4$ symmetry globally, while $h_i$ is a random field at each site. 
In the spirit of the mean field analysis we approximate this model by the infinite range interaction, where $J_{i,j}\rightarrow J/N$. Then mean field theory becomes exact \cite{Schneider1977}.
When $h_i=0$, we obtain the familiar equation of state for the order parameter $\phi=\langle\tau^z\rangle$ is ($\beta\equiv1/k_{B}T$)
\begin{equation}
\phi=\tanh[\beta(2J\phi+h_0)].
\end{equation}
The nematic transition temperature of the system without disorder
follows as $T_{s}^{\left(0\right)}=2J$. 

If we now allow for random stress, the corresponding mean-field equation
of state for the order parameter becomes 
\begin{align}
\overline{\phi} & =\int dhp_{\sigma}(h)\tanh\frac{2J\overline{\phi}+h_{0}+h}{k_{B}T},
\end{align}
which is solved self-consistently for $\overline{\phi}$. To find
the nematic phase transition temperature $T_{s}/J$ as a function
of disorder $\sigma/J$, we take $h_{0}=0$ and  expand the
hyperbolic tangent while keeping only terms linear in $\overline{\phi}$
to obtain 
\begin{equation}
\overline{\phi}=\frac{2J\overline{\phi}}{k_{B}T_{s}}C\left(\frac{\sigma}{k_{B}T_{s}}\right),
\end{equation}
where the function $C$ is given by
\begin{equation}
C\left(t\right)=\int dx\frac{e^{-\frac{x^{2}}{2}}}{\sqrt{2\pi}}\tanh^{2}\left(xt\right).
\end{equation}
The transition temperature $T_{s}\left(\sigma\right)$ of the system
with disorder is determined by the condition 
\begin{equation}
T_{s}=T_{s}^{\left(0\right)}C\left(\frac{\sigma}{k_{B}T_{s}}\right).
\end{equation}
 Random strain reduces the transition temperature, but the transition
itself remains sharp. In distinction, global external strain $h_{0}$ would
smear the transition. The nematic transition reaches $T_{s}=0$ when
$\sigma/J=\pi/2$. As we will see, NMR experiments for FeSe suggest
that $\sigma$ is significantly smaller than $\pi J/2$ where disorder
destroys nematic order completely.

The nematic susceptibility above $T_{s}$ is given by 
\begin{eqnarray}
\chi_{{\rm nem}}\left(T\right) & =\left.\frac{\partial\overline{\phi}}{\partial h_{0}}\right|_{h_{0}=0}=\frac{C\left(\frac{\sigma}{k_{B}T}\right)}{T-T_{s}^{\left(0\right)}C\left(\frac{\sigma}{k_{B}T}\right)}.
\end{eqnarray}
As mentioned above, we have to interpret the effective Ising model
as the one that includes all allowed couplings, including the ones
that are mediated by the lattice. In fact the mean-field treatment
of the clean system is possible because we have included this lattice
coupling which leads to an effective long-range interaction \cite{Karahasanovic2016}.
Hence, the nematic susceptibility $\chi_{{\rm nem}}$ is proportional
to the inverse elastic constant $C_{66}$ \cite{Fernandes2010}. It diverges at the actual
thermodynamic phase transition, not at the lower Curie-Weiss temperature
that one deduces from Raman or elasto-resistivity measurements \cite{Gallais2016}.

Finally, we determine the Edwards-Anderson order parameter. Taking
$h_{0}=0$ above $T_{s}$ where $\overline{\phi}=0$, we have 
\begin{align}
q_{{\rm EA}}\left(T\right) & =\int dhp_{\sigma}(h)\tanh^{2}\frac{h}{k_{B}T}\nonumber \\
 & =1-C\left(\frac{\sigma}{k_{B}T}\right).
\end{align}
Obviously, $q_{EA}$ above $T_{s}$ only depends on temperature and
the strength of random strain $\sigma$. As we will see below, this
explains the ``scaling'' of the NMR line width
as function of pressure. Changing the nematic coupling constant will
then not change the $T$ dependence of $q_{EA}$, but merely set the
temperature $T_{s}=T_{s}^{\left(0\right)}\left(1-q_{{\rm EA}}\left(T_{s}\right)\right)$
where it deviates from its high-temperature behavior. Clearly, the
finding that $q_{EA}\left(T>T_{s}\right)$ is independent of $J$
is only valid within the mean-field approximation. We can turn this
reasoning around and conclude that a temperature dependent NMR line
width that is independent of the value of $T_{s}$ strongly supports
that the inhomogeneous nematic state above $T_{s}$ is well captured within
a mean field treatment. 

We also note that the Edwards-Anderson parameter is not proportional
to the nematic susceptibility. However, both are  related via
\begin{equation}
\chi_{{\rm nem}}\left(T\right)=\frac{1-q_{{\rm EA}}\left(T\right)}{T-T_{s}^{\left(0\right)}\left(1-q_{{\rm EA}}\left(T\right)\right)}.
\end{equation}
Thus, in principle it is possible to determine the Edwards-Anderson
parameter from the nematic susceptibility. In practice, the available
accuracy of data for $\chi_{{\rm nem}}\left(T\right)$ turns out not
to be sufficient to determine $q_{{\rm EA}}\left(T\right)$. The reason is that $q_{{\rm EA}}\left(T\right)$ for FeSe  
turns out to be significantly smaller that unity. This also
explains that the $T$-dependence of $q_{{\rm EA}}$ does not significantly
change the Curie-Weiss behavior of the nematic susceptibility. Thus $\chi_{\rm nem}$ is not a sensitive indicator of inhomogeneous nematicity while local probes, such as  NMR line-width measurements,  are able to reveal
the existence and temperature dependence of inhomogeneous nematic
regions.

\begin{figure}[t]
\centering
\includegraphics[width=\columnwidth]{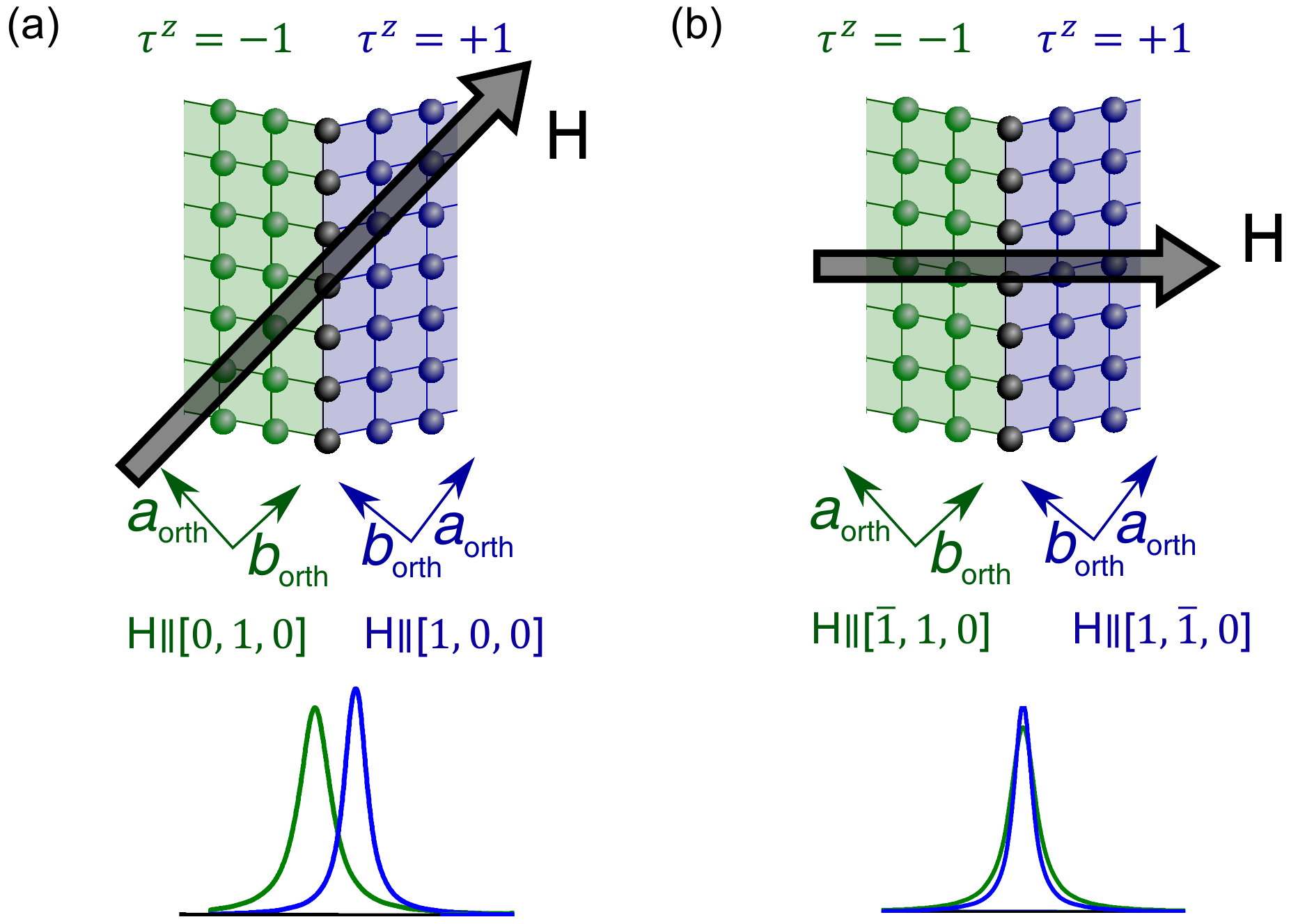}
\caption{The Ising variable $\tau^z = \pm1$ breaks the four-fold symmetry in two different ways, resulting in the formation
of domains when local nematic order is present.
In (a), a magnetic field $H$ is applied such that the nuclei in the two domains experience symmetry-inequivalent field directions.
Due to the anisotropy of magnetic susceptibility in the nematic state, the nuclei in the two domains see different local
hyperfine fields, resulting in a two-peak NMR spectrum, as illustrated schematically. In contrast, when the field is rotated 45$^\circ$ as in (b), the nuclei experience symmetry equivalent hyperfine fields, and the NMR spectrum appears as a single peak. 
}
\label{fig:IsingNematic}
\end{figure}

\begin{figure*}[t]
\centering
\includegraphics[width=\textwidth]{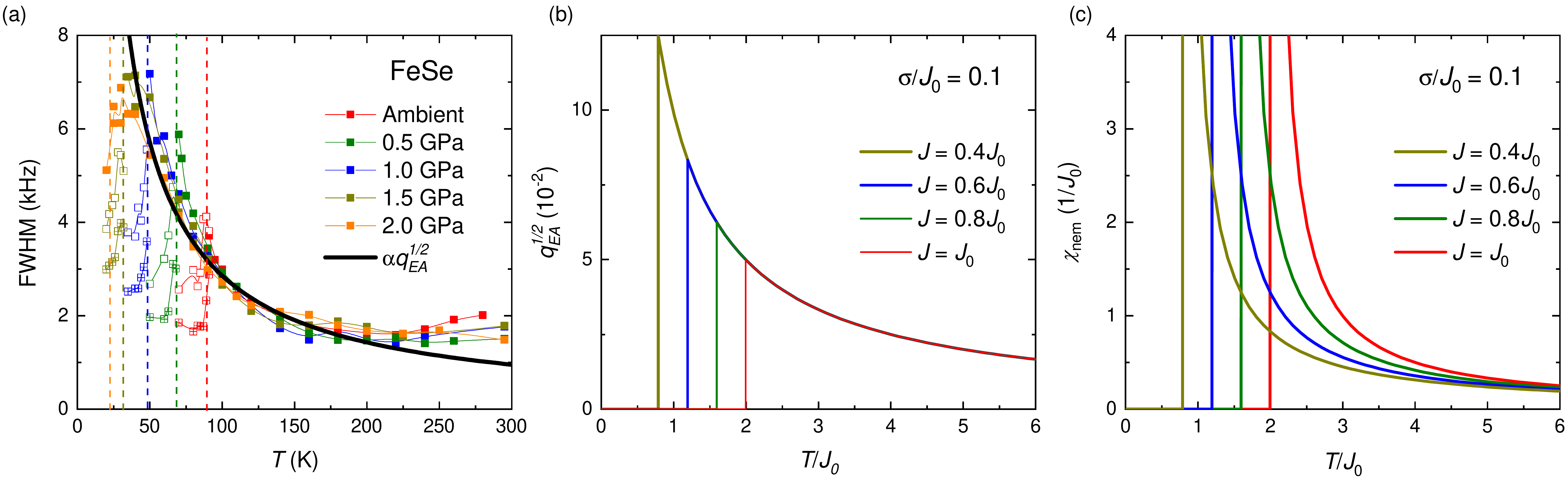}
\caption{(a) FWHM of $^{77}$Se NMR spectrum in FeSe under indicated hydrostatic pressures \cite{Wiecki2017}. Overlaid is the scaled Edwards-Anderson parameter $\alpha q_{EA}^{1/2}$ for disorder strength $\sigma/J=0.1$
and $T_s=30$ K.
Vertical dashed lines denote $T_s$. Note that at 2.0 GPa $T_s$ becomes a joint nematic-magnetic transition \cite{Kothapalli_2016}. 
Below $T_s$, the FWHM of the low-frequency NMR peak (open symbols) is greater than that of the high-frequency peak (hatched symbols).
(b) Square root of the Edwards-Anderson parameter $q_{EA}^{1/2}$ for decreasing values of nematic coupling $J$ at constant disorder strength $\sigma$. $J_0$ corresponds to the strength of the nematic coupling at ambient pressure, while pressure reduces $J$ below $J_0$. Notice that $q_{\rm EA}$ above $T_s$ is independent of this nematic interaction $J$.
(c) Nematic susceptibility $\chi_{\rm nem}$ for decreasing values of nematic coupling $J$ at constant disorder strength $\sigma$.
The pressure independent behavior of the NMR FWHM in the PM state (a) resembles the Edwards-Anderson parameter $q_{EA}^{1/2}$ (b) and not the
nematic susceptibility $\chi_{\rm nem}$ (c), as anticipated theoretically in Eq. (\ref{eqn:FWHMqEA}).
}
\label{fig:dataqEAchi}
\end{figure*}
\section{Summary of NMR Results}

Since $^{77}$Se is a $I=1/2$ nucleus, there are no quadrupole satellite lines or quadrupole shifts of the spectrum. Therefore, in the high-temperature tetragonal state, the NMR spectrum is a single peak. In the nematic state below $T_s$, there are nematic-twin domains. 
As illustrated in Fig. \ref{fig:IsingNematic}(a), half
of the domains experience $H\|a_O$ ($H\|[100]_O$) and the other half experience $H\|b_O$ ($H\|[010]_O$), where $a_O>b_O$. 
Due to the anisotropic magnetic susceptibility of the nematic phase \cite{He2018,Zhou_2020}, the hyperfine field at the nucleus will be different in the two domains, causing the NMR spectrum to split into two peaks in the nematic ordered state.
If the field is instead applied along $H\|[110]_O$ (Fig. \ref{fig:IsingNematic}(b)), then both types of domain see a symmetry-equivalent magnetic field, and there is only a single NMR peak in the nematic state, even though there are still domains. 
Recent NMR measurements under mechanical strain have revealed that the higher frequency NMR peak comes 
from the domains which experience $H\|a_O$, while the lower frequency peak comes from the $H\|b_O$ domains \cite{Zhou_2020}.

In the tetragonal phase, the FWHM of the single NMR peak increases on cooling towards the nematic state. However, this broadening is observed only when $H\|[100]_O$ and not when 
$H\|[110]_O$ \cite{Wiecki2017,Wang2017}. This observation provides clear evidence that the broadening is related to nematicity and implies the existence of a short-range nematic domain structure in the tetragonal state of FeSe. 
Since this effect is observed in the NMR spectrum, the fluctuating nematic domain structure is static on the time scale set by the inverse NMR linewidth $\sim1/(1 {\rm kHz})$.

\section{Comparison of Theory and Experiment}

\begin{figure}[t]
\centering
\includegraphics[width=0.8\columnwidth]{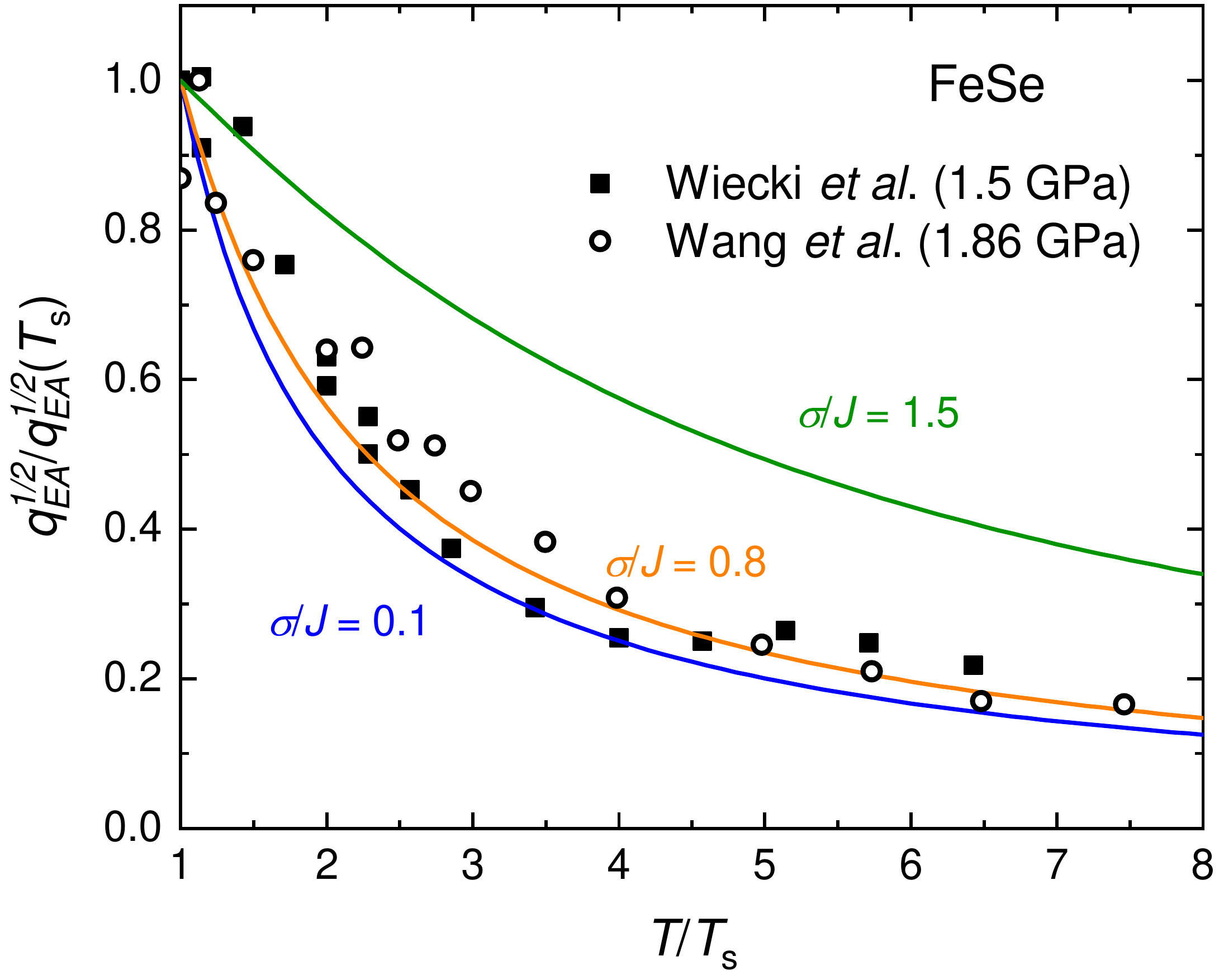}
\caption{NMR FWHM in FeSe normalized at $T_s$ as a function of $T/T_s$ at indicated hydrostatic pressures \cite{Wiecki2017,Wang2017}. 
The data are compared to $q_{EA}^{1/2}/q_{EA}^{1/2}(T_s)$ calculated in the random-field Ising nematic model at the mean-field level. The only 
adjustable parameter is the disorder strength $\sigma/J$. For $\sigma/J>\pi/2$, there is no long-range nematic order.
}
\label{fig:qEAcompare}
\end{figure}

For a direct comparison between theory and experiment FeSe offers several clear advantages over
other nematic iron-based superconductors. First, FeSe has no bulk magnetic phase at ambient pressure, and thus the pure
Ising-nematic model is expected to be relevant in this system. Secondly, the $^{77}$Se nucleus has $I=1/2$ so that no complications
arise from nuclear quadrupole couplings. Finally, no dopants are present, which can introduce additional lines into the NMR spectrum \cite{Iye2015,Ning2010}.

Fig. \ref{fig:dataqEAchi}(a) shows the NMR FWHM data for FeSe under pressure. The FWHM increases on cooling towards $T_s$.
The data at different pressures follow a universal curve, simply interrupted at the appropriate 
nematic transition temperatures $T_s(p)$. 
Below $T_s(p)$ in the nematic state, the FWHM of each of the two NMR peaks is shown separately.
The FWHM of the low-frequency NMR peak (open symbols) is greater than that of the high-frequency peak (hatched symbols). 
The different FWHM of the two NMR peaks in the nematic state of FeSe has been noted in independent measurements and
is currently not understood \cite{Zhou_2020}.

Within the Ising-nematic model, the experimental data under pressure is expected to correspond to a constant disorder strength $\sigma$ (determined only by the disorder of the particular crystal), but decreasing nematic coupling $J$, starting from a value $J_0$ that corresponds to  ambient pressure.
Figs. \ref{fig:dataqEAchi}(b),(c) show the mean-field behavior of the Edwards-Anderson parameter $q_{EA}^{1/2}$ and the nematic susceptibility $\chi_{\rm nem}$ under these assumptions.
As expected, the NMR data are well described by $q_{EA}^{1/2}$ and not $\chi_{\rm nem}$ (Eq. \ref{eqn:FWHMqEA}).
The universal FWHM curve followed at all pressures is seen to be the result of the constant $\sigma$. The existence of a universal curve
 does not mean that the nematic tendency is somehow independent of pressure,
as proposed in both NMR papers \cite{Wiecki2017,Wang2017}. 
Rather this reflects the fact that $q_{EA}\left(T>T_{s}\right)$ is independent of $J$ within the mean-field approximation, as discussed above. 
Therefore, the NMR data are consistent with the Raman data, showing a suppression of nematic fluctuations 
(Fig. \ref{fig:dataqEAchi}(c)) under pressure \cite{Massat2018}. 

The FWHM data in Fig. \ref{fig:dataqEAchi}(a)
show a deviation from $q_{EA}^{1/2}$ at high temperature. This deviation is easily understood. In a usual PM state, 
the broadening reflects the spatial distribution of $K$ values due to sample inhomogeneity, and
the temperature dependence of the broadening typically follows
the temperature dependence of the magnetic susceptibility $\chi_{\rm mag}$.
In FeSe, both $\chi_{\rm mag}$~\cite{B_hmer_2017} and the NMR shift \cite{Wiecki2017,Bohmer2015,Baek_2014} increase with warming, and therefore the NMR spectrum is expected to broaden on warming.  
At high temperatures ($T \gtrsim 200~K$), the  nematic broadening described by $q_{EA}$ is small and the NMR FWHM is dominated by standard inhomogeneous broadening that increases with increasing temperature. At lower temperatures, the short-range nematic domain structure becomes the dominant source of broadening and the FWHM increases upon decreasing temperature \cite{Wiecki2017}.

At 2 GPa, the FWHM data appear to level off just above the $T_s$. At this pressure, the $T_s$ is a joint structural-magnetic transition \cite{Kothapalli_2016}.
Here, the simple Ising-nematic model may no longer be applicable when the ground state is no longer pure nematic.

A more detailed comparison of theory and experiment is shown in Fig. \ref{fig:qEAcompare}, where we plot the FWHM (normalized at $T_s$) as a function of
reduced temperature $T/T_s$ for both NMR studies \cite{Wiecki2017,Wang2017}.
We note that the FWHM at $T_s$ differs in the two studies due to different amounts of disorder in the two crystals. 
Here, we plot the data at high pressure so that $T_s$ is reduced and the data can be compared with the Ising-nematic model over the largest temperature range. 
The only  adjustable parameter in the model is the disorder strength $\sigma/J$. 
Here, the NMR data are compared with $\sigma/J=0.1$, $0.8$ and $1.5$.
For $\sigma/J>\pi/2$, there is no long-range nematic order due to the strong disorder. 
The NMR data agree well with the theoretical curves at small disorder parameters, as expected.  

\section{Conclusions}
FeSe features a local, orthorhombic nematic order in its high-temperature, nominally tetragonal phase. 
The observed agreement between the NMR line broadening and the Edwards-Anderson parameter of the disordered, random-field Ising-nematic model implies that this 
local nematic order is primarily nucleated by crystal defects. The origin of the local orthorhombicity was not explicitly considered 
in the PDF studies of FeSe \cite{Koch2019,Frandsen2019}.
We note that this short-range ordered nematicity in the high-temperature tetragonal state is  distinct from the quantum Griffiths phase recently proposed in Fe(Se,S) under pressure, where rare but large regions of local nematic droplets undergo quantum fluctuations \cite{reiss2021signatures}. In this context it is important to keep in mind that in metallic systems such quantum fluctuations are known to be strongly suppressed  due to the  coupling to the particle-hole continuum \cite{Millis2001,Millis2002,Vojta2003}, a behavior specific to Ising degrees of freedoms \cite{Vojta2005}. In fact it is this suppression of quantum fluctuations that is consistent with the effectively classical description of local nematic order  of our approach. 
Finally, our results demonstrate that the nematic fluctuations in FeSe are suppressed by hydrostatic pressure, consistent with Raman studies \cite{Massat2018},
despite the pressure-independence of the nematic broadening of the NMR lines. 
This understanding could be further confirmed by elastoresistance measurements inside a pressure cell \cite{Gati_2020}.

We thank B. M. Andersen, S.  A.  Kivelson, R. M. Fernandes and Y. Furukawa for valuable discussions. 
We acknowledge support by the Helmholtz Association under Contract No. VH-NG-1242.
This work was also supported by the German Research Foundation (DFG) under CRC/TRR 288 (Projects A02 and B01). Work in Grenoble was supported by the Laboratoire d’Excellence LANEF (ANR-10-LABX-51-01) and by the French Agence Nationale de la Recherche (ANR) under reference ANR-19-CE30-0019 (Neptun).

\bibliography{FeSeRandomFieldPaper.bbl}

\begin{thebibliography}{79}%
\makeatletter
\providecommand \@ifxundefined [1]{%
 \@ifx{#1\undefined}
}%
\providecommand \@ifnum [1]{%
 \ifnum #1\expandafter \@firstoftwo
 \else \expandafter \@secondoftwo
 \fi
}%
\providecommand \@ifx [1]{%
 \ifx #1\expandafter \@firstoftwo
 \else \expandafter \@secondoftwo
 \fi
}%
\providecommand \natexlab [1]{#1}%
\providecommand \enquote  [1]{``#1''}%
\providecommand \bibnamefont  [1]{#1}%
\providecommand \bibfnamefont [1]{#1}%
\providecommand \citenamefont [1]{#1}%
\providecommand \href@noop [0]{\@secondoftwo}%
\providecommand \href [0]{\begingroup \@sanitize@url \@href}%
\providecommand \@href[1]{\@@startlink{#1}\@@href}%
\providecommand \@@href[1]{\endgroup#1\@@endlink}%
\providecommand \@sanitize@url [0]{\catcode `\\12\catcode `\$12\catcode
  `\&12\catcode `\#12\catcode `\^12\catcode `\_12\catcode `\%12\relax}%
\providecommand \@@startlink[1]{}%
\providecommand \@@endlink[0]{}%
\providecommand \url  [0]{\begingroup\@sanitize@url \@url }%
\providecommand \@url [1]{\endgroup\@href {#1}{\urlprefix }}%
\providecommand \urlprefix  [0]{URL }%
\providecommand \Eprint [0]{\href }%
\providecommand \doibase [0]{http://dx.doi.org/}%
\providecommand \selectlanguage [0]{\@gobble}%
\providecommand \bibinfo  [0]{\@secondoftwo}%
\providecommand \bibfield  [0]{\@secondoftwo}%
\providecommand \translation [1]{[#1]}%
\providecommand \BibitemOpen [0]{}%
\providecommand \bibitemStop [0]{}%
\providecommand \bibitemNoStop [0]{.\EOS\space}%
\providecommand \EOS [0]{\spacefactor3000\relax}%
\providecommand \BibitemShut  [1]{\csname bibitem#1\endcsname}%
\let\auto@bib@innerbib\@empty
%</preamble>
\bibitem [{\citenamefont {Urbano}\ \emph {et~al.}(2007)\citenamefont {Urbano},
  \citenamefont {Young}, \citenamefont {Curro}, \citenamefont {Thompson},
  \citenamefont {Pham},\ and\ \citenamefont {Fisk}}]{Urbano2007}%
  \BibitemOpen
  \bibfield  {author} {\bibinfo {author} {\bibfnamefont {R.~R.}\ \bibnamefont
  {Urbano}}, \bibinfo {author} {\bibfnamefont {B.-L.}\ \bibnamefont {Young}},
  \bibinfo {author} {\bibfnamefont {N.~J.}\ \bibnamefont {Curro}}, \bibinfo
  {author} {\bibfnamefont {J.~D.}\ \bibnamefont {Thompson}}, \bibinfo {author}
  {\bibfnamefont {L.~D.}\ \bibnamefont {Pham}}, \ and\ \bibinfo {author}
  {\bibfnamefont {Z.}~\bibnamefont {Fisk}},\ }\bibfield  {title} {\enquote
  {\bibinfo {title} {Interacting antiferromagnetic droplets in quantum critical
  {${\mathrm{CeCoIn}}_{5}$}},}\ }\href {\doibase 10.1103/PhysRevLett.99.146402}
  {\bibfield  {journal} {\bibinfo  {journal} {Phys. Rev. Lett.}\ }\textbf
  {\bibinfo {volume} {99}},\ \bibinfo {pages} {146402} (\bibinfo {year}
  {2007})}\BibitemShut {NoStop}%
\bibitem [{\citenamefont {Seo}\ \emph {et~al.}(2013)\citenamefont {Seo},
  \citenamefont {Lu}, \citenamefont {Zhu}, \citenamefont {Urbano},
  \citenamefont {Curro}, \citenamefont {Bauer}, \citenamefont {Sidorov},
  \citenamefont {Pham}, \citenamefont {Park}, \citenamefont {Fisk},\ and\
  \citenamefont {Thompson}}]{Seo_2013}%
  \BibitemOpen
  \bibfield  {author} {\bibinfo {author} {\bibfnamefont {S.}~\bibnamefont
  {Seo}}, \bibinfo {author} {\bibfnamefont {Xin}\ \bibnamefont {Lu}}, \bibinfo
  {author} {\bibfnamefont {J-X.}\ \bibnamefont {Zhu}}, \bibinfo {author}
  {\bibfnamefont {R.~R.}\ \bibnamefont {Urbano}}, \bibinfo {author}
  {\bibfnamefont {N.}~\bibnamefont {Curro}}, \bibinfo {author} {\bibfnamefont
  {E.~D.}\ \bibnamefont {Bauer}}, \bibinfo {author} {\bibfnamefont {V.~A.}\
  \bibnamefont {Sidorov}}, \bibinfo {author} {\bibfnamefont {L.~D.}\
  \bibnamefont {Pham}}, \bibinfo {author} {\bibfnamefont {Tuson}\ \bibnamefont
  {Park}}, \bibinfo {author} {\bibfnamefont {Z.}~\bibnamefont {Fisk}}, \ and\
  \bibinfo {author} {\bibfnamefont {J.~D.}\ \bibnamefont {Thompson}},\
  }\bibfield  {title} {\enquote {\bibinfo {title} {Disorder in quantum critical
  superconductors},}\ }\href {\doibase 10.1038/nphys2820} {\bibfield  {journal}
  {\bibinfo  {journal} {Nature Physics}\ }\textbf {\bibinfo {volume} {10}},\
  \bibinfo {pages} {120--125} (\bibinfo {year} {2013})}\BibitemShut {NoStop}%
\bibitem [{\citenamefont {Julien}\ \emph {et~al.}(2000)\citenamefont {Julien},
  \citenamefont {Feh\'er}, \citenamefont {Horvati\ifmmode~\acute{c}\else
  \'{c}\fi{}}, \citenamefont {Berthier}, \citenamefont {Bakharev},
  \citenamefont {S\'egransan}, \citenamefont {Collin},\ and\ \citenamefont
  {Marucco}}]{Julien_2000}%
  \BibitemOpen
  \bibfield  {author} {\bibinfo {author} {\bibfnamefont {M.-H.}\ \bibnamefont
  {Julien}}, \bibinfo {author} {\bibfnamefont {T.}~\bibnamefont {Feh\'er}},
  \bibinfo {author} {\bibfnamefont {M.}~\bibnamefont
  {Horvati\ifmmode~\acute{c}\else \'{c}\fi{}}}, \bibinfo {author}
  {\bibfnamefont {C.}~\bibnamefont {Berthier}}, \bibinfo {author}
  {\bibfnamefont {O.~N.}\ \bibnamefont {Bakharev}}, \bibinfo {author}
  {\bibfnamefont {P.}~\bibnamefont {S\'egransan}}, \bibinfo {author}
  {\bibfnamefont {G.}~\bibnamefont {Collin}}, \ and\ \bibinfo {author}
  {\bibfnamefont {J.-F.}\ \bibnamefont {Marucco}},\ }\bibfield  {title}
  {\enquote {\bibinfo {title} {${}^{63}\mathrm{Cu}$ {NMR} evidence for enhanced
  antiferromagnetic correlations around {Zn} impurities in
  {${\mathrm{YBa}}_{2}{\mathrm{Cu}}_{3}{O}_{6.7}$}},}\ }\href {\doibase
  10.1103/PhysRevLett.84.3422} {\bibfield  {journal} {\bibinfo  {journal}
  {Phys. Rev. Lett.}\ }\textbf {\bibinfo {volume} {84}},\ \bibinfo {pages}
  {3422--3425} (\bibinfo {year} {2000})}\BibitemShut {NoStop}%
\bibitem [{\citenamefont {Kimura}\ \emph {et~al.}(2003)\citenamefont {Kimura},
  \citenamefont {Kofu}, \citenamefont {Matsumoto},\ and\ \citenamefont
  {Hirota}}]{Kimura2003}%
  \BibitemOpen
  \bibfield  {author} {\bibinfo {author} {\bibfnamefont {H.}~\bibnamefont
  {Kimura}}, \bibinfo {author} {\bibfnamefont {M.}~\bibnamefont {Kofu}},
  \bibinfo {author} {\bibfnamefont {Y.}~\bibnamefont {Matsumoto}}, \ and\
  \bibinfo {author} {\bibfnamefont {K.}~\bibnamefont {Hirota}},\ }\bibfield
  {title} {\enquote {\bibinfo {title} {Novel in-gap spin state in zn-doped
  {${\mathrm{L}\mathrm{a}}_{1.85}{\mathrm{S}\mathrm{r}}_{0.15}{\mathrm{C}\mathrm{u}\mathrm{O}}_{4}$}},}\
  }\href {\doibase 10.1103/PhysRevLett.91.067002} {\bibfield  {journal}
  {\bibinfo  {journal} {Phys. Rev. Lett.}\ }\textbf {\bibinfo {volume} {91}},\
  \bibinfo {pages} {067002} (\bibinfo {year} {2003})}\BibitemShut {NoStop}%
\bibitem [{\citenamefont {Berthier}\ \emph {et~al.}(1978)\citenamefont
  {Berthier}, \citenamefont {Jerome},\ and\ \citenamefont
  {Molinie}}]{Berthier_1978}%
  \BibitemOpen
  \bibfield  {author} {\bibinfo {author} {\bibfnamefont {C.}~\bibnamefont
  {Berthier}}, \bibinfo {author} {\bibfnamefont {D.}~\bibnamefont {Jerome}}, \
  and\ \bibinfo {author} {\bibfnamefont {C.}~\bibnamefont {Molinie}},\
  }\bibfield  {title} {\enquote {\bibinfo {title} {{NMR} study on a
  2h-{NbSe}$_2$ single crystal: A microscopic investigation of the charge
  density waves state},}\ }\href@noop {} {\bibfield  {journal} {\bibinfo
  {journal} {Journal of Physics C: Solid State Physics}\ }\textbf {\bibinfo
  {volume} {11}},\ \bibinfo {pages} {797--814} (\bibinfo {year}
  {1978})}\BibitemShut {NoStop}%
\bibitem [{\citenamefont {Arguello}\ \emph {et~al.}(2014)\citenamefont
  {Arguello}, \citenamefont {Chockalingam}, \citenamefont {Rosenthal},
  \citenamefont {Zhao}, \citenamefont {Guti\'errez}, \citenamefont {Kang},
  \citenamefont {Chung}, \citenamefont {Fernandes}, \citenamefont {Jia},
  \citenamefont {Millis}, \citenamefont {Cava},\ and\ \citenamefont
  {Pasupathy}}]{Arguello2014}%
  \BibitemOpen
  \bibfield  {author} {\bibinfo {author} {\bibfnamefont {C.~J.}\ \bibnamefont
  {Arguello}}, \bibinfo {author} {\bibfnamefont {S.~P.}\ \bibnamefont
  {Chockalingam}}, \bibinfo {author} {\bibfnamefont {E.~P.}\ \bibnamefont
  {Rosenthal}}, \bibinfo {author} {\bibfnamefont {L.}~\bibnamefont {Zhao}},
  \bibinfo {author} {\bibfnamefont {C.}~\bibnamefont {Guti\'errez}}, \bibinfo
  {author} {\bibfnamefont {J.~H.}\ \bibnamefont {Kang}}, \bibinfo {author}
  {\bibfnamefont {W.~C.}\ \bibnamefont {Chung}}, \bibinfo {author}
  {\bibfnamefont {R.~M.}\ \bibnamefont {Fernandes}}, \bibinfo {author}
  {\bibfnamefont {S.}~\bibnamefont {Jia}}, \bibinfo {author} {\bibfnamefont
  {A.~J.}\ \bibnamefont {Millis}}, \bibinfo {author} {\bibfnamefont {R.~J.}\
  \bibnamefont {Cava}}, \ and\ \bibinfo {author} {\bibfnamefont {A.~N.}\
  \bibnamefont {Pasupathy}},\ }\bibfield  {title} {\enquote {\bibinfo {title}
  {Visualizing the charge density wave transition in {$2H$-${\text{NbSe}}_{2}$}
  in real space},}\ }\href {\doibase 10.1103/PhysRevB.89.235115} {\bibfield
  {journal} {\bibinfo  {journal} {Phys. Rev. B}\ }\textbf {\bibinfo {volume}
  {89}},\ \bibinfo {pages} {235115} (\bibinfo {year} {2014})}\BibitemShut
  {NoStop}%
\bibitem [{\citenamefont {Chatterjee}\ \emph {et~al.}(2015)\citenamefont
  {Chatterjee}, \citenamefont {Zhao}, \citenamefont {Iavarone}, \citenamefont
  {Di~Capua}, \citenamefont {Castellan}, \citenamefont {Karapetrov},
  \citenamefont {Malliakas}, \citenamefont {Kanatzidis}, \citenamefont {Claus},
  \citenamefont {Ruff}, \citenamefont {Weber}, \citenamefont {van Wezel},
  \citenamefont {Campuzano}, \citenamefont {Osborn}, \citenamefont {Randeria},
  \citenamefont {Trivedi}, \citenamefont {Norman},\ and\ \citenamefont
  {Rosenkranz}}]{Chatterjee_2015}%
  \BibitemOpen
  \bibfield  {author} {\bibinfo {author} {\bibfnamefont {U.}~\bibnamefont
  {Chatterjee}}, \bibinfo {author} {\bibfnamefont {J.}~\bibnamefont {Zhao}},
  \bibinfo {author} {\bibfnamefont {M.}~\bibnamefont {Iavarone}}, \bibinfo
  {author} {\bibfnamefont {R.}~\bibnamefont {Di~Capua}}, \bibinfo {author}
  {\bibfnamefont {J.~P.}\ \bibnamefont {Castellan}}, \bibinfo {author}
  {\bibfnamefont {G.}~\bibnamefont {Karapetrov}}, \bibinfo {author}
  {\bibfnamefont {C.~D.}\ \bibnamefont {Malliakas}}, \bibinfo {author}
  {\bibfnamefont {M.~G.}\ \bibnamefont {Kanatzidis}}, \bibinfo {author}
  {\bibfnamefont {H.}~\bibnamefont {Claus}}, \bibinfo {author} {\bibfnamefont
  {J.~P.~C.}\ \bibnamefont {Ruff}}, \bibinfo {author} {\bibfnamefont
  {F.}~\bibnamefont {Weber}}, \bibinfo {author} {\bibfnamefont
  {J.}~\bibnamefont {van Wezel}}, \bibinfo {author} {\bibfnamefont {J.~C.}\
  \bibnamefont {Campuzano}}, \bibinfo {author} {\bibfnamefont {R.}~\bibnamefont
  {Osborn}}, \bibinfo {author} {\bibfnamefont {M.}~\bibnamefont {Randeria}},
  \bibinfo {author} {\bibfnamefont {N.}~\bibnamefont {Trivedi}}, \bibinfo
  {author} {\bibfnamefont {M.~R.}\ \bibnamefont {Norman}}, \ and\ \bibinfo
  {author} {\bibfnamefont {S.}~\bibnamefont {Rosenkranz}},\ }\bibfield  {title}
  {\enquote {\bibinfo {title} {Emergence of coherence in the charge-density
  wave state of {2H-NbSe$_2$}},}\ }\href@noop {} {\bibfield  {journal}
  {\bibinfo  {journal} {Nature Communications}\ }\textbf {\bibinfo {volume}
  {6}},\ \bibinfo {pages} {6313} (\bibinfo {year} {2015})}\BibitemShut
  {NoStop}%
\bibitem [{\citenamefont {Yue}\ \emph {et~al.}(2020)\citenamefont {Yue},
  \citenamefont {Xue}, \citenamefont {Li}, \citenamefont {Hu}, \citenamefont
  {Barbour}, \citenamefont {Zheng}, \citenamefont {Wang}, \citenamefont {Feng},
  \citenamefont {Wilkins}, \citenamefont {Mazzoli}, \citenamefont {Comin},\
  and\ \citenamefont {Li}}]{Yue_2020}%
  \BibitemOpen
  \bibfield  {author} {\bibinfo {author} {\bibfnamefont {Li}~\bibnamefont
  {Yue}}, \bibinfo {author} {\bibfnamefont {Shangjie}\ \bibnamefont {Xue}},
  \bibinfo {author} {\bibfnamefont {Jiarui}\ \bibnamefont {Li}}, \bibinfo
  {author} {\bibfnamefont {Wen}\ \bibnamefont {Hu}}, \bibinfo {author}
  {\bibfnamefont {Andi}\ \bibnamefont {Barbour}}, \bibinfo {author}
  {\bibfnamefont {Feipeng}\ \bibnamefont {Zheng}}, \bibinfo {author}
  {\bibfnamefont {Lichen}\ \bibnamefont {Wang}}, \bibinfo {author}
  {\bibfnamefont {Ji}~\bibnamefont {Feng}}, \bibinfo {author} {\bibfnamefont
  {Stuart~B.}\ \bibnamefont {Wilkins}}, \bibinfo {author} {\bibfnamefont
  {Claudio}\ \bibnamefont {Mazzoli}}, \bibinfo {author} {\bibfnamefont
  {Riccardo}\ \bibnamefont {Comin}}, \ and\ \bibinfo {author} {\bibfnamefont
  {Yuan}\ \bibnamefont {Li}},\ }\bibfield  {title} {\enquote {\bibinfo {title}
  {Distinction between pristine and disorder-perturbed charge density waves in
  {ZrTe$_3$}},}\ }\href {\doibase 10.1038/s41467-019-13813-y} {\bibfield
  {journal} {\bibinfo  {journal} {Nature Communications}\ }\textbf {\bibinfo
  {volume} {11}},\ \bibinfo {pages} {98} (\bibinfo {year} {2020})}\BibitemShut
  {NoStop}%
\bibitem [{\citenamefont {Liu}\ \emph {et~al.}(2021)\citenamefont {Liu},
  \citenamefont {Zhu}, \citenamefont {Liu}, \citenamefont {Deng}, \citenamefont
  {Zhou}, \citenamefont {Li}, \citenamefont {Sun}, \citenamefont {Huang},
  \citenamefont {Li}, \citenamefont {Du}, \citenamefont {Wang}, \citenamefont
  {Guan}, \citenamefont {Mao}, \citenamefont {Sui}, \citenamefont {Wu},
  \citenamefont {Yin}, \citenamefont {Cheng},\ and\ \citenamefont
  {Pan}}]{liu_2021}%
  \BibitemOpen
  \bibfield  {author} {\bibinfo {author} {\bibfnamefont {Limin}\ \bibnamefont
  {Liu}}, \bibinfo {author} {\bibfnamefont {Changjiang}\ \bibnamefont {Zhu}},
  \bibinfo {author} {\bibfnamefont {Z.~Y.}\ \bibnamefont {Liu}}, \bibinfo
  {author} {\bibfnamefont {Hanbin}\ \bibnamefont {Deng}}, \bibinfo {author}
  {\bibfnamefont {X.~B.}\ \bibnamefont {Zhou}}, \bibinfo {author}
  {\bibfnamefont {Yuan}\ \bibnamefont {Li}}, \bibinfo {author} {\bibfnamefont
  {Yingkai}\ \bibnamefont {Sun}}, \bibinfo {author} {\bibfnamefont {Xiong}\
  \bibnamefont {Huang}}, \bibinfo {author} {\bibfnamefont {Shuaishuai}\
  \bibnamefont {Li}}, \bibinfo {author} {\bibfnamefont {Xin}\ \bibnamefont
  {Du}}, \bibinfo {author} {\bibfnamefont {Zheng}\ \bibnamefont {Wang}},
  \bibinfo {author} {\bibfnamefont {Tong}\ \bibnamefont {Guan}}, \bibinfo
  {author} {\bibfnamefont {Hanqing}\ \bibnamefont {Mao}}, \bibinfo {author}
  {\bibfnamefont {Y.}~\bibnamefont {Sui}}, \bibinfo {author} {\bibfnamefont
  {Rui}\ \bibnamefont {Wu}}, \bibinfo {author} {\bibfnamefont {Jia-Xin}\
  \bibnamefont {Yin}}, \bibinfo {author} {\bibfnamefont {J.~G.}\ \bibnamefont
  {Cheng}}, \ and\ \bibinfo {author} {\bibfnamefont {Shuheng~H.}\ \bibnamefont
  {Pan}},\ }\href@noop {} {\enquote {\bibinfo {title} {Thermal dynamics of
  charge density wave pinning in {ZrTe$_3$}},}\ } (\bibinfo {year} {2021}),\
  \Eprint {http://arxiv.org/abs/2105.02640} {arXiv:2105.02640
  [cond-mat.str-el]} \BibitemShut {NoStop}%
\bibitem [{\citenamefont {Wu}\ \emph {et~al.}(2015)\citenamefont {Wu},
  \citenamefont {Mayaffre}, \citenamefont {Krämer}, \citenamefont
  {Horvati{\'{c}}}, \citenamefont {Berthier}, \citenamefont {Hardy},
  \citenamefont {Liang}, \citenamefont {Bonn},\ and\ \citenamefont
  {Julien}}]{Wu_2015}%
  \BibitemOpen
  \bibfield  {author} {\bibinfo {author} {\bibfnamefont {Tao}\ \bibnamefont
  {Wu}}, \bibinfo {author} {\bibfnamefont {Hadrien}\ \bibnamefont {Mayaffre}},
  \bibinfo {author} {\bibfnamefont {Steffen}\ \bibnamefont {Krämer}}, \bibinfo
  {author} {\bibfnamefont {Mladen}\ \bibnamefont {Horvati{\'{c}}}}, \bibinfo
  {author} {\bibfnamefont {Claude}\ \bibnamefont {Berthier}}, \bibinfo {author}
  {\bibfnamefont {W.N.}\ \bibnamefont {Hardy}}, \bibinfo {author}
  {\bibfnamefont {Ruixing}\ \bibnamefont {Liang}}, \bibinfo {author}
  {\bibfnamefont {D.A.}\ \bibnamefont {Bonn}}, \ and\ \bibinfo {author}
  {\bibfnamefont {Marc-Henri}\ \bibnamefont {Julien}},\ }\bibfield  {title}
  {\enquote {\bibinfo {title} {Incipient charge order observed by {NMR} in the
  normal state of {YBa$_2$Cu$_3$O$_y$}},}\ }\href {\doibase 10.1038/ncomms7438}
  {\bibfield  {journal} {\bibinfo  {journal} {Nature Communications}\ }\textbf
  {\bibinfo {volume} {6}} (\bibinfo {year} {2015}),\
  10.1038/ncomms7438}\BibitemShut {NoStop}%
\bibitem [{\citenamefont {Weitering}\ \emph {et~al.}(1999)\citenamefont
  {Weitering}, \citenamefont {Carpinelli}, \citenamefont {Melechko},
  \citenamefont {Zhang}, \citenamefont {Bartkowiak},\ and\ \citenamefont
  {Plummer}}]{Weitering2107}%
  \BibitemOpen
  \bibfield  {author} {\bibinfo {author} {\bibfnamefont {Hanno~H.}\
  \bibnamefont {Weitering}}, \bibinfo {author} {\bibfnamefont {Joseph~M.}\
  \bibnamefont {Carpinelli}}, \bibinfo {author} {\bibfnamefont {Anatoli~V.}\
  \bibnamefont {Melechko}}, \bibinfo {author} {\bibfnamefont {Jiandi}\
  \bibnamefont {Zhang}}, \bibinfo {author} {\bibfnamefont {Miroslaw}\
  \bibnamefont {Bartkowiak}}, \ and\ \bibinfo {author} {\bibfnamefont
  {E.~Ward}\ \bibnamefont {Plummer}},\ }\bibfield  {title} {\enquote {\bibinfo
  {title} {Defect-mediated condensation of a charge density wave},}\ }\href
  {\doibase 10.1126/science.285.5436.2107} {\bibfield  {journal} {\bibinfo
  {journal} {Science}\ }\textbf {\bibinfo {volume} {285}},\ \bibinfo {pages}
  {2107--2110} (\bibinfo {year} {1999})},\ \Eprint
  {http://arxiv.org/abs/https://science.sciencemag.org/content/285/5436/2107.full.pdf}
  {https://science.sciencemag.org/content/285/5436/2107.full.pdf} \BibitemShut
  {NoStop}%
\bibitem [{\citenamefont {Kasahara}\ \emph {et~al.}(2012)\citenamefont
  {Kasahara}, \citenamefont {Shi}, \citenamefont {Hashimoto}, \citenamefont
  {Tonegawa}, \citenamefont {Mizukami}, \citenamefont {Shibauchi},
  \citenamefont {Sugimoto}, \citenamefont {Fukuda}, \citenamefont {Terashima},
  \citenamefont {Nevidomskyy},\ and\ \citenamefont {Matsuda}}]{Kasahara2012}%
  \BibitemOpen
  \bibfield  {author} {\bibinfo {author} {\bibfnamefont {S.}~\bibnamefont
  {Kasahara}}, \bibinfo {author} {\bibfnamefont {H.~J.}\ \bibnamefont {Shi}},
  \bibinfo {author} {\bibfnamefont {K.}~\bibnamefont {Hashimoto}}, \bibinfo
  {author} {\bibfnamefont {S.}~\bibnamefont {Tonegawa}}, \bibinfo {author}
  {\bibfnamefont {Y.}~\bibnamefont {Mizukami}}, \bibinfo {author}
  {\bibfnamefont {T.}~\bibnamefont {Shibauchi}}, \bibinfo {author}
  {\bibfnamefont {K.}~\bibnamefont {Sugimoto}}, \bibinfo {author}
  {\bibfnamefont {T.}~\bibnamefont {Fukuda}}, \bibinfo {author} {\bibfnamefont
  {T.}~\bibnamefont {Terashima}}, \bibinfo {author} {\bibfnamefont {Andriy~H.}\
  \bibnamefont {Nevidomskyy}}, \ and\ \bibinfo {author} {\bibfnamefont
  {Y.}~\bibnamefont {Matsuda}},\ }\bibfield  {title} {\enquote {\bibinfo
  {title} {Electronic nematicity above the structural and superconducting
  transition in {BaFe$_2$(As$_{1-x}$P$_x$)$_2$}},}\ }\href {\doibase
  10.1038/nature11178} {\bibfield  {journal} {\bibinfo  {journal} {Nature}\
  }\textbf {\bibinfo {volume} {486}},\ \bibinfo {pages} {382--385} (\bibinfo
  {year} {2012})}\BibitemShut {NoStop}%
\bibitem [{\citenamefont {Yi}\ \emph {et~al.}(2011)\citenamefont {Yi},
  \citenamefont {Lu}, \citenamefont {Chu}, \citenamefont {Analytis},
  \citenamefont {Sorini}, \citenamefont {Kemper}, \citenamefont {Moritz},
  \citenamefont {Mo}, \citenamefont {Moore}, \citenamefont {Hashimoto},
  \citenamefont {Lee}, \citenamefont {Hussain}, \citenamefont {Devereaux},
  \citenamefont {Fisher},\ and\ \citenamefont {Shen}}]{Yi6878}%
  \BibitemOpen
  \bibfield  {author} {\bibinfo {author} {\bibfnamefont {Ming}\ \bibnamefont
  {Yi}}, \bibinfo {author} {\bibfnamefont {Donghui}\ \bibnamefont {Lu}},
  \bibinfo {author} {\bibfnamefont {Jiun-Haw}\ \bibnamefont {Chu}}, \bibinfo
  {author} {\bibfnamefont {James~G.}\ \bibnamefont {Analytis}}, \bibinfo
  {author} {\bibfnamefont {Adam~P.}\ \bibnamefont {Sorini}}, \bibinfo {author}
  {\bibfnamefont {Alexander~F.}\ \bibnamefont {Kemper}}, \bibinfo {author}
  {\bibfnamefont {Brian}\ \bibnamefont {Moritz}}, \bibinfo {author}
  {\bibfnamefont {Sung-Kwan}\ \bibnamefont {Mo}}, \bibinfo {author}
  {\bibfnamefont {Rob~G.}\ \bibnamefont {Moore}}, \bibinfo {author}
  {\bibfnamefont {Makoto}\ \bibnamefont {Hashimoto}}, \bibinfo {author}
  {\bibfnamefont {Wei-Sheng}\ \bibnamefont {Lee}}, \bibinfo {author}
  {\bibfnamefont {Zahid}\ \bibnamefont {Hussain}}, \bibinfo {author}
  {\bibfnamefont {Thomas~P.}\ \bibnamefont {Devereaux}}, \bibinfo {author}
  {\bibfnamefont {Ian~R.}\ \bibnamefont {Fisher}}, \ and\ \bibinfo {author}
  {\bibfnamefont {Zhi-Xun}\ \bibnamefont {Shen}},\ }\bibfield  {title}
  {\enquote {\bibinfo {title} {Symmetry-breaking orbital anisotropy observed
  for detwinned {Ba(Fe$_{1-x}$Co$_x$)$_2$As$_2$} above the spin density wave
  transition},}\ }\href {\doibase 10.1073/pnas.1015572108} {\bibfield
  {journal} {\bibinfo  {journal} {Proceedings of the National Academy of
  Sciences}\ }\textbf {\bibinfo {volume} {108}},\ \bibinfo {pages} {6878--6883}
  (\bibinfo {year} {2011})},\ \Eprint
  {http://arxiv.org/abs/https://www.pnas.org/content/108/17/6878.full.pdf}
  {https://www.pnas.org/content/108/17/6878.full.pdf} \BibitemShut {NoStop}%
\bibitem [{\citenamefont {Wang}\ \emph {et~al.}(2018)\citenamefont {Wang},
  \citenamefont {Song}, \citenamefont {Cao}, \citenamefont {Tseng},
  \citenamefont {Keller}, \citenamefont {Li}, \citenamefont {Harriger},
  \citenamefont {Tian}, \citenamefont {Chi}, \citenamefont {Yu}, \citenamefont
  {Nevidomskyy},\ and\ \citenamefont {Dai}}]{Wang2018}%
  \BibitemOpen
  \bibfield  {author} {\bibinfo {author} {\bibfnamefont {Weiyi}\ \bibnamefont
  {Wang}}, \bibinfo {author} {\bibfnamefont {Yu}~\bibnamefont {Song}}, \bibinfo
  {author} {\bibfnamefont {Chongde}\ \bibnamefont {Cao}}, \bibinfo {author}
  {\bibfnamefont {Kuo-Feng}\ \bibnamefont {Tseng}}, \bibinfo {author}
  {\bibfnamefont {Thomas}\ \bibnamefont {Keller}}, \bibinfo {author}
  {\bibfnamefont {Yu}~\bibnamefont {Li}}, \bibinfo {author} {\bibfnamefont
  {L.~W.}\ \bibnamefont {Harriger}}, \bibinfo {author} {\bibfnamefont {Wei}\
  \bibnamefont {Tian}}, \bibinfo {author} {\bibfnamefont {Songxue}\
  \bibnamefont {Chi}}, \bibinfo {author} {\bibfnamefont {Rong}\ \bibnamefont
  {Yu}}, \bibinfo {author} {\bibfnamefont {Andriy~H.}\ \bibnamefont
  {Nevidomskyy}}, \ and\ \bibinfo {author} {\bibfnamefont {Pengcheng}\
  \bibnamefont {Dai}},\ }\bibfield  {title} {\enquote {\bibinfo {title} {Local
  orthorhombic lattice distortions in the paramagnetic tetragonal phase of
  superconducting {NaFe$_{1-x}$Ni$_x$As}},}\ }\href {\doibase
  10.1038/s41467-018-05529-2} {\bibfield  {journal} {\bibinfo  {journal}
  {Nature Communications}\ }\textbf {\bibinfo {volume} {9}} (\bibinfo {year}
  {2018}),\ 10.1038/s41467-018-05529-2}\BibitemShut {NoStop}%
\bibitem [{\citenamefont {Hong}\ \emph {et~al.}(2020)\citenamefont {Hong},
  \citenamefont {Caglieris}, \citenamefont {Kappenberger}, \citenamefont
  {Wurmehl}, \citenamefont {Aswartham}, \citenamefont {Scaravaggi},
  \citenamefont {Lepucki}, \citenamefont {Wolter}, \citenamefont {Grafe},
  \citenamefont {B\"uchner},\ and\ \citenamefont {Hess}}]{Hong2020}%
  \BibitemOpen
  \bibfield  {author} {\bibinfo {author} {\bibfnamefont {Xiaochen}\
  \bibnamefont {Hong}}, \bibinfo {author} {\bibfnamefont {Federico}\
  \bibnamefont {Caglieris}}, \bibinfo {author} {\bibfnamefont {Rhea}\
  \bibnamefont {Kappenberger}}, \bibinfo {author} {\bibfnamefont {Sabine}\
  \bibnamefont {Wurmehl}}, \bibinfo {author} {\bibfnamefont {Saicharan}\
  \bibnamefont {Aswartham}}, \bibinfo {author} {\bibfnamefont {Francesco}\
  \bibnamefont {Scaravaggi}}, \bibinfo {author} {\bibfnamefont {Piotr}\
  \bibnamefont {Lepucki}}, \bibinfo {author} {\bibfnamefont {Anja U.~B.}\
  \bibnamefont {Wolter}}, \bibinfo {author} {\bibfnamefont {Hans-Joachim}\
  \bibnamefont {Grafe}}, \bibinfo {author} {\bibfnamefont {Bernd}\ \bibnamefont
  {B\"uchner}}, \ and\ \bibinfo {author} {\bibfnamefont {Christian}\
  \bibnamefont {Hess}},\ }\bibfield  {title} {\enquote {\bibinfo {title}
  {Evolution of the nematic susceptibility in
  {${\mathrm{LaFe}}_{1\ensuremath{-}x}{\mathrm{Co}}_{x}\mathrm{AsO}$}},}\
  }\href {\doibase 10.1103/PhysRevLett.125.067001} {\bibfield  {journal}
  {\bibinfo  {journal} {Phys. Rev. Lett.}\ }\textbf {\bibinfo {volume} {125}},\
  \bibinfo {pages} {067001} (\bibinfo {year} {2020})}\BibitemShut {NoStop}%
\bibitem [{\citenamefont {Rosenthal}\ \emph {et~al.}(2014)\citenamefont
  {Rosenthal}, \citenamefont {Andrade}, \citenamefont {Arguello}, \citenamefont
  {Fernandes}, \citenamefont {Xing}, \citenamefont {Wang}, \citenamefont {Jin},
  \citenamefont {Millis},\ and\ \citenamefont {Pasupathy}}]{Rosenthal_2014}%
  \BibitemOpen
  \bibfield  {author} {\bibinfo {author} {\bibfnamefont {E.~P.}\ \bibnamefont
  {Rosenthal}}, \bibinfo {author} {\bibfnamefont {E.~F.}\ \bibnamefont
  {Andrade}}, \bibinfo {author} {\bibfnamefont {C.~J.}\ \bibnamefont
  {Arguello}}, \bibinfo {author} {\bibfnamefont {R.~M.}\ \bibnamefont
  {Fernandes}}, \bibinfo {author} {\bibfnamefont {L.~Y.}\ \bibnamefont {Xing}},
  \bibinfo {author} {\bibfnamefont {X.~C.}\ \bibnamefont {Wang}}, \bibinfo
  {author} {\bibfnamefont {C.~Q.}\ \bibnamefont {Jin}}, \bibinfo {author}
  {\bibfnamefont {A.~J.}\ \bibnamefont {Millis}}, \ and\ \bibinfo {author}
  {\bibfnamefont {A.~N.}\ \bibnamefont {Pasupathy}},\ }\bibfield  {title}
  {\enquote {\bibinfo {title} {Visualization of electron nematicity and
  unidirectional antiferroic fluctuations at high temperatures in {NaFeAs}},}\
  }\href {\doibase 10.1038/nphys2870} {\bibfield  {journal} {\bibinfo
  {journal} {Nature Physics}\ }\textbf {\bibinfo {volume} {10}},\ \bibinfo
  {pages} {225--232} (\bibinfo {year} {2014})}\BibitemShut {NoStop}%
\bibitem [{\citenamefont {Iye}\ \emph {et~al.}(2015)\citenamefont {Iye},
  \citenamefont {Julien}, \citenamefont {Mayaffre}, \citenamefont
  {Horvati{\'{c}}}, \citenamefont {Berthier}, \citenamefont {Ishida},
  \citenamefont {Ikeda}, \citenamefont {Kasahara}, \citenamefont {Shibauchi},\
  and\ \citenamefont {Matsuda}}]{Iye2015}%
  \BibitemOpen
  \bibfield  {author} {\bibinfo {author} {\bibfnamefont {Tetsuya}\ \bibnamefont
  {Iye}}, \bibinfo {author} {\bibfnamefont {Marc-Henri}\ \bibnamefont
  {Julien}}, \bibinfo {author} {\bibfnamefont {Hadrien}\ \bibnamefont
  {Mayaffre}}, \bibinfo {author} {\bibfnamefont {Mladen}\ \bibnamefont
  {Horvati{\'{c}}}}, \bibinfo {author} {\bibfnamefont {Claude}\ \bibnamefont
  {Berthier}}, \bibinfo {author} {\bibfnamefont {Kenji}\ \bibnamefont
  {Ishida}}, \bibinfo {author} {\bibfnamefont {Hiroaki}\ \bibnamefont {Ikeda}},
  \bibinfo {author} {\bibfnamefont {Shigeru}\ \bibnamefont {Kasahara}},
  \bibinfo {author} {\bibfnamefont {Takasada}\ \bibnamefont {Shibauchi}}, \
  and\ \bibinfo {author} {\bibfnamefont {Yuji}\ \bibnamefont {Matsuda}},\
  }\bibfield  {title} {\enquote {\bibinfo {title} {Emergence of orbital
  nematicity in the tetragonal phase of {BaFe$_2$(As$_{1-x}$P$_x$)$_2$}},}\
  }\href {\doibase 10.7566/jpsj.84.043705} {\bibfield  {journal} {\bibinfo
  {journal} {Journal of the Physical Society of Japan}\ }\textbf {\bibinfo
  {volume} {84}},\ \bibinfo {pages} {043705} (\bibinfo {year}
  {2015})}\BibitemShut {NoStop}%
\bibitem [{\citenamefont {Zhou}\ \emph {et~al.}(2016)\citenamefont {Zhou},
  \citenamefont {Xing}, \citenamefont {Wang}, \citenamefont {Jin},\ and\
  \citenamefont {Zheng}}]{Zhou2016}%
  \BibitemOpen
  \bibfield  {author} {\bibinfo {author} {\bibfnamefont {R.}~\bibnamefont
  {Zhou}}, \bibinfo {author} {\bibfnamefont {L.~Y.}\ \bibnamefont {Xing}},
  \bibinfo {author} {\bibfnamefont {X.~C.}\ \bibnamefont {Wang}}, \bibinfo
  {author} {\bibfnamefont {C.~Q.}\ \bibnamefont {Jin}}, \ and\ \bibinfo
  {author} {\bibfnamefont {Guo-qing}\ \bibnamefont {Zheng}},\ }\bibfield
  {title} {\enquote {\bibinfo {title} {Orbital order and spin nematicity in the
  tetragonal phase of the electron-doped iron pnictides
  {${\mathrm{NaFe}}_{1\ensuremath{-}x}{\mathrm{Co}}_{x}\mathrm{As}$}},}\ }\href
  {\doibase 10.1103/PhysRevB.93.060502} {\bibfield  {journal} {\bibinfo
  {journal} {Phys. Rev. B}\ }\textbf {\bibinfo {volume} {93}},\ \bibinfo
  {pages} {060502} (\bibinfo {year} {2016})}\BibitemShut {NoStop}%
\bibitem [{\citenamefont {Toyoda}\ \emph
  {et~al.}(2018{\natexlab{a}})\citenamefont {Toyoda}, \citenamefont
  {Kobayashi},\ and\ \citenamefont {Itoh}}]{Toyoda2018}%
  \BibitemOpen
  \bibfield  {author} {\bibinfo {author} {\bibfnamefont {Masayuki}\
  \bibnamefont {Toyoda}}, \bibinfo {author} {\bibfnamefont {Yoshiaki}\
  \bibnamefont {Kobayashi}}, \ and\ \bibinfo {author} {\bibfnamefont
  {Masayuki}\ \bibnamefont {Itoh}},\ }\bibfield  {title} {\enquote {\bibinfo
  {title} {Nematic fluctuations in iron arsenides {NaFeAs} and {LiFeAs} probed
  by {$^{75}\mathrm{As}$} {NMR}},}\ }\href {\doibase
  10.1103/PhysRevB.97.094515} {\bibfield  {journal} {\bibinfo  {journal} {Phys.
  Rev. B}\ }\textbf {\bibinfo {volume} {97}},\ \bibinfo {pages} {094515}
  (\bibinfo {year} {2018}{\natexlab{a}})}\BibitemShut {NoStop}%
\bibitem [{\citenamefont {Toyoda}\ \emph
  {et~al.}(2018{\natexlab{b}})\citenamefont {Toyoda}, \citenamefont {Ichikawa},
  \citenamefont {Kobayashi}, \citenamefont {Sato},\ and\ \citenamefont
  {Itoh}}]{Toyoda_2018b}%
  \BibitemOpen
  \bibfield  {author} {\bibinfo {author} {\bibfnamefont {Masayuki}\
  \bibnamefont {Toyoda}}, \bibinfo {author} {\bibfnamefont {Akihiro}\
  \bibnamefont {Ichikawa}}, \bibinfo {author} {\bibfnamefont {Yoshiaki}\
  \bibnamefont {Kobayashi}}, \bibinfo {author} {\bibfnamefont {Masatoshi}\
  \bibnamefont {Sato}}, \ and\ \bibinfo {author} {\bibfnamefont {Masayuki}\
  \bibnamefont {Itoh}},\ }\bibfield  {title} {\enquote {\bibinfo {title}
  {In-plane anisotropy of the electric field gradient in
  {${\mathrm{Ba}(\mathrm{Fe}}_{1\ensuremath{-}x}{\mathrm{Co}}_{x}{)}_{2}{\mathrm{As}}_{2}$}
  observed by {$^{75}\mathrm{As}$} {NMR}},}\ }\href {\doibase
  10.1103/PhysRevB.97.174507} {\bibfield  {journal} {\bibinfo  {journal} {Phys.
  Rev. B}\ }\textbf {\bibinfo {volume} {97}},\ \bibinfo {pages} {174507}
  (\bibinfo {year} {2018}{\natexlab{b}})}\BibitemShut {NoStop}%
\bibitem [{\citenamefont {Dioguardi}\ \emph {et~al.}(2015)\citenamefont
  {Dioguardi}, \citenamefont {Lawson}, \citenamefont {Bush}, \citenamefont
  {Crocker}, \citenamefont {Shirer}, \citenamefont {Nisson}, \citenamefont
  {Kissikov}, \citenamefont {Ran}, \citenamefont {Bud'ko}, \citenamefont
  {Canfield}, \citenamefont {Yuan}, \citenamefont {Kuhns}, \citenamefont
  {Reyes}, \citenamefont {Grafe},\ and\ \citenamefont {Curro}}]{Dioguardi2015}%
  \BibitemOpen
  \bibfield  {author} {\bibinfo {author} {\bibfnamefont {A.~P.}\ \bibnamefont
  {Dioguardi}}, \bibinfo {author} {\bibfnamefont {M.~M.}\ \bibnamefont
  {Lawson}}, \bibinfo {author} {\bibfnamefont {B.~T.}\ \bibnamefont {Bush}},
  \bibinfo {author} {\bibfnamefont {J.}~\bibnamefont {Crocker}}, \bibinfo
  {author} {\bibfnamefont {K.~R.}\ \bibnamefont {Shirer}}, \bibinfo {author}
  {\bibfnamefont {D.~M.}\ \bibnamefont {Nisson}}, \bibinfo {author}
  {\bibfnamefont {T.}~\bibnamefont {Kissikov}}, \bibinfo {author}
  {\bibfnamefont {S.}~\bibnamefont {Ran}}, \bibinfo {author} {\bibfnamefont
  {S.~L.}\ \bibnamefont {Bud'ko}}, \bibinfo {author} {\bibfnamefont {P.~C.}\
  \bibnamefont {Canfield}}, \bibinfo {author} {\bibfnamefont {S.}~\bibnamefont
  {Yuan}}, \bibinfo {author} {\bibfnamefont {P.~L.}\ \bibnamefont {Kuhns}},
  \bibinfo {author} {\bibfnamefont {A.~P.}\ \bibnamefont {Reyes}}, \bibinfo
  {author} {\bibfnamefont {H.-J.}\ \bibnamefont {Grafe}}, \ and\ \bibinfo
  {author} {\bibfnamefont {N.~J.}\ \bibnamefont {Curro}},\ }\bibfield  {title}
  {\enquote {\bibinfo {title} {{NMR} evidence for inhomogeneous glassy behavior
  driven by nematic fluctuations in iron arsenide superconductors},}\ }\href
  {\doibase 10.1103/PhysRevB.92.165116} {\bibfield  {journal} {\bibinfo
  {journal} {Phys. Rev. B}\ }\textbf {\bibinfo {volume} {92}},\ \bibinfo
  {pages} {165116} (\bibinfo {year} {2015})}\BibitemShut {NoStop}%
\bibitem [{\citenamefont {Dioguardi}\ \emph {et~al.}(2016)\citenamefont
  {Dioguardi}, \citenamefont {Kissikov}, \citenamefont {Lin}, \citenamefont
  {Shirer}, \citenamefont {Lawson}, \citenamefont {Grafe}, \citenamefont {Chu},
  \citenamefont {Fisher}, \citenamefont {Fernandes},\ and\ \citenamefont
  {Curro}}]{Dioguardi2016}%
  \BibitemOpen
  \bibfield  {author} {\bibinfo {author} {\bibfnamefont {A.~P.}\ \bibnamefont
  {Dioguardi}}, \bibinfo {author} {\bibfnamefont {T.}~\bibnamefont {Kissikov}},
  \bibinfo {author} {\bibfnamefont {C.~H.}\ \bibnamefont {Lin}}, \bibinfo
  {author} {\bibfnamefont {K.~R.}\ \bibnamefont {Shirer}}, \bibinfo {author}
  {\bibfnamefont {M.~M.}\ \bibnamefont {Lawson}}, \bibinfo {author}
  {\bibfnamefont {H.-J.}\ \bibnamefont {Grafe}}, \bibinfo {author}
  {\bibfnamefont {J.-H.}\ \bibnamefont {Chu}}, \bibinfo {author} {\bibfnamefont
  {I.~R.}\ \bibnamefont {Fisher}}, \bibinfo {author} {\bibfnamefont {R.~M.}\
  \bibnamefont {Fernandes}}, \ and\ \bibinfo {author} {\bibfnamefont {N.~J.}\
  \bibnamefont {Curro}},\ }\bibfield  {title} {\enquote {\bibinfo {title} {{NMR
  Evidence for Inhomogeneous Nematic Fluctuations in
  ${\mathrm{BaFe}}_{2}({\mathrm{As}}_{1\ensuremath{-}x}{\mathrm{P}}_{x}{)}_{2}$}},}\
  }\href {\doibase 10.1103/PhysRevLett.116.107202} {\bibfield  {journal}
  {\bibinfo  {journal} {Phys. Rev. Lett.}\ }\textbf {\bibinfo {volume} {116}},\
  \bibinfo {pages} {107202} (\bibinfo {year} {2016})}\BibitemShut {NoStop}%
\bibitem [{\citenamefont {Bossoni}\ \emph {et~al.}(2013)\citenamefont
  {Bossoni}, \citenamefont {Carretta}, \citenamefont {Halperin}, \citenamefont
  {Oh}, \citenamefont {Reyes}, \citenamefont {Kuhns},\ and\ \citenamefont
  {Canfield}}]{Bossoni2013}%
  \BibitemOpen
  \bibfield  {author} {\bibinfo {author} {\bibfnamefont {L.}~\bibnamefont
  {Bossoni}}, \bibinfo {author} {\bibfnamefont {P.}~\bibnamefont {Carretta}},
  \bibinfo {author} {\bibfnamefont {W.~P.}\ \bibnamefont {Halperin}}, \bibinfo
  {author} {\bibfnamefont {S.}~\bibnamefont {Oh}}, \bibinfo {author}
  {\bibfnamefont {A.}~\bibnamefont {Reyes}}, \bibinfo {author} {\bibfnamefont
  {P.}~\bibnamefont {Kuhns}}, \ and\ \bibinfo {author} {\bibfnamefont {P.~C.}\
  \bibnamefont {Canfield}},\ }\bibfield  {title} {\enquote {\bibinfo {title}
  {Evidence of unconventional low-frequency dynamics in the normal phase of
  {Ba(Fe${}_{1\ensuremath{-}x}$Rh${}_{x}$)${}_{2}$As${}_{2}$} iron-based
  superconductors},}\ }\href {\doibase 10.1103/PhysRevB.88.100503} {\bibfield
  {journal} {\bibinfo  {journal} {Phys. Rev. B}\ }\textbf {\bibinfo {volume}
  {88}},\ \bibinfo {pages} {100503} (\bibinfo {year} {2013})}\BibitemShut
  {NoStop}%
\bibitem [{\citenamefont {Bossoni}\ \emph {et~al.}(2016)\citenamefont
  {Bossoni}, \citenamefont {Moroni}, \citenamefont {Julien}, \citenamefont
  {Mayaffre}, \citenamefont {Canfield}, \citenamefont {Reyes}, \citenamefont
  {Halperin},\ and\ \citenamefont {Carretta}}]{Bossoni2016}%
  \BibitemOpen
  \bibfield  {author} {\bibinfo {author} {\bibfnamefont {L.}~\bibnamefont
  {Bossoni}}, \bibinfo {author} {\bibfnamefont {M.}~\bibnamefont {Moroni}},
  \bibinfo {author} {\bibfnamefont {M.~H.}\ \bibnamefont {Julien}}, \bibinfo
  {author} {\bibfnamefont {H.}~\bibnamefont {Mayaffre}}, \bibinfo {author}
  {\bibfnamefont {P.~C.}\ \bibnamefont {Canfield}}, \bibinfo {author}
  {\bibfnamefont {A.}~\bibnamefont {Reyes}}, \bibinfo {author} {\bibfnamefont
  {W.~P.}\ \bibnamefont {Halperin}}, \ and\ \bibinfo {author} {\bibfnamefont
  {P.}~\bibnamefont {Carretta}},\ }\bibfield  {title} {\enquote {\bibinfo
  {title} {Persistence of slow fluctuations in the overdoped regime of
  {$\mathrm{Ba}({\mathrm{Fe}}_{1\ensuremath{-}x}{\mathrm{Rh}}_{x}){}_{2}{\mathrm{As}}_{2}$}
  superconductors},}\ }\href {\doibase 10.1103/PhysRevB.93.224517} {\bibfield
  {journal} {\bibinfo  {journal} {Phys. Rev. B}\ }\textbf {\bibinfo {volume}
  {93}},\ \bibinfo {pages} {224517} (\bibinfo {year} {2016})}\BibitemShut
  {NoStop}%
\bibitem [{\citenamefont {Frandsen}\ \emph {et~al.}(2018)\citenamefont
  {Frandsen}, \citenamefont {Taddei}, \citenamefont {Bugaris}, \citenamefont
  {Stadel}, \citenamefont {Yi}, \citenamefont {Acharya}, \citenamefont
  {Osborn}, \citenamefont {Rosenkranz}, \citenamefont {Chmaissem},\ and\
  \citenamefont {Birgeneau}}]{Frandsen2018}%
  \BibitemOpen
  \bibfield  {author} {\bibinfo {author} {\bibfnamefont {Benjamin~A.}\
  \bibnamefont {Frandsen}}, \bibinfo {author} {\bibfnamefont {Keith~M.}\
  \bibnamefont {Taddei}}, \bibinfo {author} {\bibfnamefont {Daniel~E.}\
  \bibnamefont {Bugaris}}, \bibinfo {author} {\bibfnamefont {Ryan}\
  \bibnamefont {Stadel}}, \bibinfo {author} {\bibfnamefont {Ming}\ \bibnamefont
  {Yi}}, \bibinfo {author} {\bibfnamefont {Arani}\ \bibnamefont {Acharya}},
  \bibinfo {author} {\bibfnamefont {Raymond}\ \bibnamefont {Osborn}}, \bibinfo
  {author} {\bibfnamefont {Stephan}\ \bibnamefont {Rosenkranz}}, \bibinfo
  {author} {\bibfnamefont {Omar}\ \bibnamefont {Chmaissem}}, \ and\ \bibinfo
  {author} {\bibfnamefont {Robert~J.}\ \bibnamefont {Birgeneau}},\ }\bibfield
  {title} {\enquote {\bibinfo {title} {Widespread orthorhombic fluctuations in
  the {$(\mathrm{Sr},\mathrm{Na}){\mathrm{Fe}}_{2}{\mathrm{As}}_{2}$} family of
  superconductors},}\ }\href {\doibase 10.1103/PhysRevB.98.180505} {\bibfield
  {journal} {\bibinfo  {journal} {Phys. Rev. B}\ }\textbf {\bibinfo {volume}
  {98}},\ \bibinfo {pages} {180505} (\bibinfo {year} {2018})}\BibitemShut
  {NoStop}%
\bibitem [{\citenamefont {Frandsen}\ \emph {et~al.}(2017)\citenamefont
  {Frandsen}, \citenamefont {Taddei}, \citenamefont {Yi}, \citenamefont
  {Frano}, \citenamefont {Guguchia}, \citenamefont {Yu}, \citenamefont {Si},
  \citenamefont {Bugaris}, \citenamefont {Stadel}, \citenamefont {Osborn},
  \citenamefont {Rosenkranz}, \citenamefont {Chmaissem},\ and\ \citenamefont
  {Birgeneau}}]{Frandsen2017}%
  \BibitemOpen
  \bibfield  {author} {\bibinfo {author} {\bibfnamefont {Benjamin~A.}\
  \bibnamefont {Frandsen}}, \bibinfo {author} {\bibfnamefont {Keith~M.}\
  \bibnamefont {Taddei}}, \bibinfo {author} {\bibfnamefont {Ming}\ \bibnamefont
  {Yi}}, \bibinfo {author} {\bibfnamefont {Alex}\ \bibnamefont {Frano}},
  \bibinfo {author} {\bibfnamefont {Zurab}\ \bibnamefont {Guguchia}}, \bibinfo
  {author} {\bibfnamefont {Rong}\ \bibnamefont {Yu}}, \bibinfo {author}
  {\bibfnamefont {Qimiao}\ \bibnamefont {Si}}, \bibinfo {author} {\bibfnamefont
  {Daniel~E.}\ \bibnamefont {Bugaris}}, \bibinfo {author} {\bibfnamefont
  {Ryan}\ \bibnamefont {Stadel}}, \bibinfo {author} {\bibfnamefont {Raymond}\
  \bibnamefont {Osborn}}, \bibinfo {author} {\bibfnamefont {Stephan}\
  \bibnamefont {Rosenkranz}}, \bibinfo {author} {\bibfnamefont {Omar}\
  \bibnamefont {Chmaissem}}, \ and\ \bibinfo {author} {\bibfnamefont
  {Robert~J.}\ \bibnamefont {Birgeneau}},\ }\bibfield  {title} {\enquote
  {\bibinfo {title} {Local orthorhombicity in the magnetic ${C}_{4}$ phase of
  the hole-doped iron-arsenide superconductor
  {${\mathrm{Sr}}_{1\ensuremath{-}x}{\mathrm{Na}}_{x}{\mathrm{Fe}}_{2}{\mathrm{As}}_{2}$}},}\
  }\href {\doibase 10.1103/PhysRevLett.119.187001} {\bibfield  {journal}
  {\bibinfo  {journal} {Phys. Rev. Lett.}\ }\textbf {\bibinfo {volume} {119}},\
  \bibinfo {pages} {187001} (\bibinfo {year} {2017})}\BibitemShut {NoStop}%
\bibitem [{\citenamefont {Wu}\ \emph {et~al.}(2021)\citenamefont {Wu},
  \citenamefont {Song}, \citenamefont {He}, \citenamefont {Frano},
  \citenamefont {Yi}, \citenamefont {Chen}, \citenamefont {Uchiyama},
  \citenamefont {Alatas}, \citenamefont {Said}, \citenamefont {Wang},
  \citenamefont {Wolf}, \citenamefont {Meingast},\ and\ \citenamefont
  {Birgeneau}}]{Wu2021}%
  \BibitemOpen
  \bibfield  {author} {\bibinfo {author} {\bibfnamefont {Shan}\ \bibnamefont
  {Wu}}, \bibinfo {author} {\bibfnamefont {Yu}~\bibnamefont {Song}}, \bibinfo
  {author} {\bibfnamefont {Yu}~\bibnamefont {He}}, \bibinfo {author}
  {\bibfnamefont {Alex}\ \bibnamefont {Frano}}, \bibinfo {author}
  {\bibfnamefont {Ming}\ \bibnamefont {Yi}}, \bibinfo {author} {\bibfnamefont
  {Xiang}\ \bibnamefont {Chen}}, \bibinfo {author} {\bibfnamefont {Hiroshi}\
  \bibnamefont {Uchiyama}}, \bibinfo {author} {\bibfnamefont {Ahmet}\
  \bibnamefont {Alatas}}, \bibinfo {author} {\bibfnamefont {Ayman~H.}\
  \bibnamefont {Said}}, \bibinfo {author} {\bibfnamefont {Liran}\ \bibnamefont
  {Wang}}, \bibinfo {author} {\bibfnamefont {Thomas}\ \bibnamefont {Wolf}},
  \bibinfo {author} {\bibfnamefont {Christoph}\ \bibnamefont {Meingast}}, \
  and\ \bibinfo {author} {\bibfnamefont {Robert~J.}\ \bibnamefont
  {Birgeneau}},\ }\bibfield  {title} {\enquote {\bibinfo {title} {Short-range
  nematic fluctuations in
  {${\mathrm{Sr}}_{1\ensuremath{-}x}{\mathrm{Na}}_{x}{\mathrm{Fe}}_{2}{\mathrm{As}}_{2}$}
  superconductors},}\ }\href {\doibase 10.1103/PhysRevLett.126.107001}
  {\bibfield  {journal} {\bibinfo  {journal} {Phys. Rev. Lett.}\ }\textbf
  {\bibinfo {volume} {126}},\ \bibinfo {pages} {107001} (\bibinfo {year}
  {2021})}\BibitemShut {NoStop}%
\bibitem [{\citenamefont {Weber}\ \emph {et~al.}(2018)\citenamefont {Weber},
  \citenamefont {Parshall}, \citenamefont {Pintschovius}, \citenamefont
  {Castellan}, \citenamefont {Kauth}, \citenamefont {Merz}, \citenamefont
  {Wolf}, \citenamefont {Sch\"utt}, \citenamefont {Schmalian}, \citenamefont
  {Fernandes},\ and\ \citenamefont {Reznik}}]{Weber2018}%
  \BibitemOpen
  \bibfield  {author} {\bibinfo {author} {\bibfnamefont {F.}~\bibnamefont
  {Weber}}, \bibinfo {author} {\bibfnamefont {D.}~\bibnamefont {Parshall}},
  \bibinfo {author} {\bibfnamefont {L.}~\bibnamefont {Pintschovius}}, \bibinfo
  {author} {\bibfnamefont {J.-P.}\ \bibnamefont {Castellan}}, \bibinfo {author}
  {\bibfnamefont {M.}~\bibnamefont {Kauth}}, \bibinfo {author} {\bibfnamefont
  {M.}~\bibnamefont {Merz}}, \bibinfo {author} {\bibfnamefont {Th.}\
  \bibnamefont {Wolf}}, \bibinfo {author} {\bibfnamefont {M.}~\bibnamefont
  {Sch\"utt}}, \bibinfo {author} {\bibfnamefont {J.}~\bibnamefont {Schmalian}},
  \bibinfo {author} {\bibfnamefont {R.~M.}\ \bibnamefont {Fernandes}}, \ and\
  \bibinfo {author} {\bibfnamefont {D.}~\bibnamefont {Reznik}},\ }\bibfield
  {title} {\enquote {\bibinfo {title} {Soft phonons reveal the nematic
  correlation length in
  {$\mathrm{Ba}{({\mathrm{Fe}}_{0.94}{\mathrm{Co}}_{0.06})}_{2}{\mathrm{As}}_{2}$}},}\
  }\href {\doibase 10.1103/PhysRevB.98.014516} {\bibfield  {journal} {\bibinfo
  {journal} {Phys. Rev. B}\ }\textbf {\bibinfo {volume} {98}},\ \bibinfo
  {pages} {014516} (\bibinfo {year} {2018})}\BibitemShut {NoStop}%
\bibitem [{\citenamefont {Merritt}\ \emph {et~al.}(2020)\citenamefont
  {Merritt}, \citenamefont {Weber}, \citenamefont {Castellan}, \citenamefont
  {Wolf}, \citenamefont {Ishikawa}, \citenamefont {Said}, \citenamefont
  {Alatas}, \citenamefont {Fernandes}, \citenamefont {Baron},\ and\
  \citenamefont {Reznik}}]{Merritt2020}%
  \BibitemOpen
  \bibfield  {author} {\bibinfo {author} {\bibfnamefont {A.~M.}\ \bibnamefont
  {Merritt}}, \bibinfo {author} {\bibfnamefont {F.}~\bibnamefont {Weber}},
  \bibinfo {author} {\bibfnamefont {J.-P.}\ \bibnamefont {Castellan}}, \bibinfo
  {author} {\bibfnamefont {Th.}\ \bibnamefont {Wolf}}, \bibinfo {author}
  {\bibfnamefont {D.}~\bibnamefont {Ishikawa}}, \bibinfo {author}
  {\bibfnamefont {A.~H.}\ \bibnamefont {Said}}, \bibinfo {author}
  {\bibfnamefont {A.}~\bibnamefont {Alatas}}, \bibinfo {author} {\bibfnamefont
  {R.~M.}\ \bibnamefont {Fernandes}}, \bibinfo {author} {\bibfnamefont
  {A.~Q.~R.}\ \bibnamefont {Baron}}, \ and\ \bibinfo {author} {\bibfnamefont
  {D.}~\bibnamefont {Reznik}},\ }\bibfield  {title} {\enquote {\bibinfo {title}
  {Nematic correlation length in iron-based superconductors probed by inelastic
  x-ray scattering},}\ }\href {\doibase 10.1103/PhysRevLett.124.157001}
  {\bibfield  {journal} {\bibinfo  {journal} {Phys. Rev. Lett.}\ }\textbf
  {\bibinfo {volume} {124}},\ \bibinfo {pages} {157001} (\bibinfo {year}
  {2020})}\BibitemShut {NoStop}%
\bibitem [{\citenamefont {Kauth}\ \emph {et~al.}(2020)\citenamefont {Kauth},
  \citenamefont {Rosenkranz}, \citenamefont {Said}, \citenamefont {Taddei},
  \citenamefont {Wolf},\ and\ \citenamefont {Weber}}]{Kauth2020}%
  \BibitemOpen
  \bibfield  {author} {\bibinfo {author} {\bibfnamefont {M.}~\bibnamefont
  {Kauth}}, \bibinfo {author} {\bibfnamefont {S.}~\bibnamefont {Rosenkranz}},
  \bibinfo {author} {\bibfnamefont {A.~H.}\ \bibnamefont {Said}}, \bibinfo
  {author} {\bibfnamefont {K.~M.}\ \bibnamefont {Taddei}}, \bibinfo {author}
  {\bibfnamefont {Th.}\ \bibnamefont {Wolf}}, \ and\ \bibinfo {author}
  {\bibfnamefont {F.}~\bibnamefont {Weber}},\ }\bibfield  {title} {\enquote
  {\bibinfo {title} {Soft elastic constants from phonon spectroscopy in
  hole-doped
  {${\mathrm{Ba}}_{1\ensuremath{-}x}{(\mathrm{K},\mathrm{Na})}_{x}{\mathrm{Fe}}_{2}{\mathrm{As}}_{2}$
  and
  ${\mathrm{Sr}}_{1\ensuremath{-}x}{\mathrm{Na}}_{x}{\mathrm{Fe}}_{2}{\mathrm{As}}_{2}$}},}\
  }\href {\doibase 10.1103/PhysRevB.102.144526} {\bibfield  {journal} {\bibinfo
   {journal} {Phys. Rev. B}\ }\textbf {\bibinfo {volume} {102}},\ \bibinfo
  {pages} {144526} (\bibinfo {year} {2020})}\BibitemShut {NoStop}%
\bibitem [{\citenamefont {Baek}\ \emph {et~al.}(2016)\citenamefont {Baek},
  \citenamefont {Efremov}, \citenamefont {Ok}, \citenamefont {Kim},
  \citenamefont {van~den Brink},\ and\ \citenamefont {B\"uchner}}]{Baek2016}%
  \BibitemOpen
  \bibfield  {author} {\bibinfo {author} {\bibfnamefont {S.-H.}\ \bibnamefont
  {Baek}}, \bibinfo {author} {\bibfnamefont {D.~V.}\ \bibnamefont {Efremov}},
  \bibinfo {author} {\bibfnamefont {J.~M.}\ \bibnamefont {Ok}}, \bibinfo
  {author} {\bibfnamefont {J.~S.}\ \bibnamefont {Kim}}, \bibinfo {author}
  {\bibfnamefont {Jeroen}\ \bibnamefont {van~den Brink}}, \ and\ \bibinfo
  {author} {\bibfnamefont {B.}~\bibnamefont {B\"uchner}},\ }\bibfield  {title}
  {\enquote {\bibinfo {title} {Nematicity and in-plane anisotropy of
  superconductivity in $\ensuremath{\beta}\ensuremath{-}\mathrm{FeSe}$ detected
  by $^{77}\mathrm{Se}$ nuclear magnetic resonance},}\ }\href {\doibase
  10.1103/PhysRevB.93.180502} {\bibfield  {journal} {\bibinfo  {journal} {Phys.
  Rev. B}\ }\textbf {\bibinfo {volume} {93}},\ \bibinfo {pages} {180502}
  (\bibinfo {year} {2016})}\BibitemShut {NoStop}%
\bibitem [{\citenamefont {Nie}\ \emph {et~al.}(2014)\citenamefont {Nie},
  \citenamefont {Tarjus},\ and\ \citenamefont {Kivelson}}]{Nie2014}%
  \BibitemOpen
  \bibfield  {author} {\bibinfo {author} {\bibfnamefont {Laimei}\ \bibnamefont
  {Nie}}, \bibinfo {author} {\bibfnamefont {Gilles}\ \bibnamefont {Tarjus}}, \
  and\ \bibinfo {author} {\bibfnamefont {Steven~Allan}\ \bibnamefont
  {Kivelson}},\ }\bibfield  {title} {\enquote {\bibinfo {title} {Quenched
  disorder and vestigial nematicity in the pseudogap regime of the cuprates},}\
  }\href {\doibase 10.1073/pnas.1406019111} {\bibfield  {journal} {\bibinfo
  {journal} {Proceedings of the National Academy of Sciences}\ }\textbf
  {\bibinfo {volume} {111}},\ \bibinfo {pages} {7980--7985} (\bibinfo {year}
  {2014})},\ \Eprint
  {http://arxiv.org/abs/https://www.pnas.org/content/111/22/7980.full.pdf}
  {https://www.pnas.org/content/111/22/7980.full.pdf} \BibitemShut {NoStop}%
\bibitem [{\citenamefont {Steffensen}\ \emph {et~al.}(2019)\citenamefont
  {Steffensen}, \citenamefont {Kotetes}, \citenamefont {Paul},\ and\
  \citenamefont {Andersen}}]{Steffensen2019}%
  \BibitemOpen
  \bibfield  {author} {\bibinfo {author} {\bibfnamefont {Daniel}\ \bibnamefont
  {Steffensen}}, \bibinfo {author} {\bibfnamefont {Panagiotis}\ \bibnamefont
  {Kotetes}}, \bibinfo {author} {\bibfnamefont {Indranil}\ \bibnamefont
  {Paul}}, \ and\ \bibinfo {author} {\bibfnamefont {Brian~M.}\ \bibnamefont
  {Andersen}},\ }\bibfield  {title} {\enquote {\bibinfo {title}
  {Disorder-induced electronic nematicity},}\ }\href {\doibase
  10.1103/PhysRevB.100.064521} {\bibfield  {journal} {\bibinfo  {journal}
  {Phys. Rev. B}\ }\textbf {\bibinfo {volume} {100}},\ \bibinfo {pages}
  {064521} (\bibinfo {year} {2019})}\BibitemShut {NoStop}%
\bibitem [{\citenamefont {Watson}\ \emph {et~al.}(2015)\citenamefont {Watson},
  \citenamefont {Kim}, \citenamefont {Haghighirad}, \citenamefont {Davies},
  \citenamefont {McCollam}, \citenamefont {Narayanan}, \citenamefont {Blake},
  \citenamefont {Chen}, \citenamefont {Ghannadzadeh}, \citenamefont
  {Schofield}, \citenamefont {Hoesch}, \citenamefont {Meingast}, \citenamefont
  {Wolf},\ and\ \citenamefont {Coldea}}]{Watson2015}%
  \BibitemOpen
  \bibfield  {author} {\bibinfo {author} {\bibfnamefont {M.~D.}\ \bibnamefont
  {Watson}}, \bibinfo {author} {\bibfnamefont {T.~K.}\ \bibnamefont {Kim}},
  \bibinfo {author} {\bibfnamefont {A.~A.}\ \bibnamefont {Haghighirad}},
  \bibinfo {author} {\bibfnamefont {N.~R.}\ \bibnamefont {Davies}}, \bibinfo
  {author} {\bibfnamefont {A.}~\bibnamefont {McCollam}}, \bibinfo {author}
  {\bibfnamefont {A.}~\bibnamefont {Narayanan}}, \bibinfo {author}
  {\bibfnamefont {S.~F.}\ \bibnamefont {Blake}}, \bibinfo {author}
  {\bibfnamefont {Y.~L.}\ \bibnamefont {Chen}}, \bibinfo {author}
  {\bibfnamefont {S.}~\bibnamefont {Ghannadzadeh}}, \bibinfo {author}
  {\bibfnamefont {A.~J.}\ \bibnamefont {Schofield}}, \bibinfo {author}
  {\bibfnamefont {M.}~\bibnamefont {Hoesch}}, \bibinfo {author} {\bibfnamefont
  {C.}~\bibnamefont {Meingast}}, \bibinfo {author} {\bibfnamefont
  {T.}~\bibnamefont {Wolf}}, \ and\ \bibinfo {author} {\bibfnamefont {A.~I.}\
  \bibnamefont {Coldea}},\ }\bibfield  {title} {\enquote {\bibinfo {title}
  {Emergence of the nematic electronic state in {FeSe}},}\ }\href {\doibase
  10.1103/PhysRevB.91.155106} {\bibfield  {journal} {\bibinfo  {journal} {Phys.
  Rev. B}\ }\textbf {\bibinfo {volume} {91}},\ \bibinfo {pages} {155106}
  (\bibinfo {year} {2015})}\BibitemShut {NoStop}%
\bibitem [{\citenamefont {B\"ohmer}\ \emph {et~al.}(2015)\citenamefont
  {B\"ohmer}, \citenamefont {Arai}, \citenamefont {Hardy}, \citenamefont
  {Hattori}, \citenamefont {Iye}, \citenamefont {Wolf}, \citenamefont
  {L\"ohneysen}, \citenamefont {Ishida},\ and\ \citenamefont
  {Meingast}}]{Bohmer2015}%
  \BibitemOpen
  \bibfield  {author} {\bibinfo {author} {\bibfnamefont {A.~E.}\ \bibnamefont
  {B\"ohmer}}, \bibinfo {author} {\bibfnamefont {T.}~\bibnamefont {Arai}},
  \bibinfo {author} {\bibfnamefont {F.}~\bibnamefont {Hardy}}, \bibinfo
  {author} {\bibfnamefont {T.}~\bibnamefont {Hattori}}, \bibinfo {author}
  {\bibfnamefont {T.}~\bibnamefont {Iye}}, \bibinfo {author} {\bibfnamefont
  {T.}~\bibnamefont {Wolf}}, \bibinfo {author} {\bibfnamefont {H.~v.}\
  \bibnamefont {L\"ohneysen}}, \bibinfo {author} {\bibfnamefont
  {K.}~\bibnamefont {Ishida}}, \ and\ \bibinfo {author} {\bibfnamefont
  {C.}~\bibnamefont {Meingast}},\ }\bibfield  {title} {\enquote {\bibinfo
  {title} {Origin of the tetragonal-to-orthorhombic phase transition in {FeSe}:
  A combined thermodynamic and {NMR} study of nematicity},}\ }\href {\doibase
  10.1103/PhysRevLett.114.027001} {\bibfield  {journal} {\bibinfo  {journal}
  {Phys. Rev. Lett.}\ }\textbf {\bibinfo {volume} {114}},\ \bibinfo {pages}
  {027001} (\bibinfo {year} {2015})}\BibitemShut {NoStop}%
\bibitem [{\citenamefont {Luo}\ \emph {et~al.}(2017)\citenamefont {Luo},
  \citenamefont {Cheng}, \citenamefont {Wang}, \citenamefont {Chiang},
  \citenamefont {Lin}, \citenamefont {Wu}, \citenamefont {Juang}, \citenamefont
  {Chareev}, \citenamefont {Volkova},\ and\ \citenamefont
  {Vasiliev}}]{Luo2017}%
  \BibitemOpen
  \bibfield  {author} {\bibinfo {author} {\bibfnamefont {Chih-Wei}\
  \bibnamefont {Luo}}, \bibinfo {author} {\bibfnamefont {Po~Chung}\
  \bibnamefont {Cheng}}, \bibinfo {author} {\bibfnamefont {Shun-Hung}\
  \bibnamefont {Wang}}, \bibinfo {author} {\bibfnamefont {Jen-Che}\
  \bibnamefont {Chiang}}, \bibinfo {author} {\bibfnamefont {Jiunn-Yuan}\
  \bibnamefont {Lin}}, \bibinfo {author} {\bibfnamefont {Kaung-Hsiung}\
  \bibnamefont {Wu}}, \bibinfo {author} {\bibfnamefont {Jenh-Yih}\ \bibnamefont
  {Juang}}, \bibinfo {author} {\bibfnamefont {Dmitry~A.}\ \bibnamefont
  {Chareev}}, \bibinfo {author} {\bibfnamefont {Olga~S.}\ \bibnamefont
  {Volkova}}, \ and\ \bibinfo {author} {\bibfnamefont {Alexander~N.}\
  \bibnamefont {Vasiliev}},\ }\bibfield  {title} {\enquote {\bibinfo {title}
  {Unveiling the hidden nematicity and spin subsystem in {FeSe}},}\ }\href
  {\doibase 10.1038/s41535-017-0036-5} {\bibfield  {journal} {\bibinfo
  {journal} {npj Quantum Materials}\ }\textbf {\bibinfo {volume} {2}} (\bibinfo
  {year} {2017}),\ 10.1038/s41535-017-0036-5}\BibitemShut {NoStop}%
\bibitem [{\citenamefont {Massat}\ \emph {et~al.}(2016)\citenamefont {Massat},
  \citenamefont {Farina}, \citenamefont {Paul}, \citenamefont {Karlsson},
  \citenamefont {Strobel}, \citenamefont {Toulemonde}, \citenamefont
  {M{\'e}asson}, \citenamefont {Cazayous}, \citenamefont {Sacuto},
  \citenamefont {Kasahara}, \citenamefont {Shibauchi}, \citenamefont
  {Matsuda},\ and\ \citenamefont {Gallais}}]{Massat2016}%
  \BibitemOpen
  \bibfield  {author} {\bibinfo {author} {\bibfnamefont {Pierre}\ \bibnamefont
  {Massat}}, \bibinfo {author} {\bibfnamefont {Donato}\ \bibnamefont {Farina}},
  \bibinfo {author} {\bibfnamefont {Indranil}\ \bibnamefont {Paul}}, \bibinfo
  {author} {\bibfnamefont {Sandra}\ \bibnamefont {Karlsson}}, \bibinfo {author}
  {\bibfnamefont {Pierre}\ \bibnamefont {Strobel}}, \bibinfo {author}
  {\bibfnamefont {Pierre}\ \bibnamefont {Toulemonde}}, \bibinfo {author}
  {\bibfnamefont {Marie-Aude}\ \bibnamefont {M{\'e}asson}}, \bibinfo {author}
  {\bibfnamefont {Maximilien}\ \bibnamefont {Cazayous}}, \bibinfo {author}
  {\bibfnamefont {Alain}\ \bibnamefont {Sacuto}}, \bibinfo {author}
  {\bibfnamefont {Shigeru}\ \bibnamefont {Kasahara}}, \bibinfo {author}
  {\bibfnamefont {Takasada}\ \bibnamefont {Shibauchi}}, \bibinfo {author}
  {\bibfnamefont {Yuji}\ \bibnamefont {Matsuda}}, \ and\ \bibinfo {author}
  {\bibfnamefont {Yann}\ \bibnamefont {Gallais}},\ }\bibfield  {title}
  {\enquote {\bibinfo {title} {Charge-induced nematicity in {FeSe}},}\ }\href
  {\doibase 10.1073/pnas.1606562113} {\bibfield  {journal} {\bibinfo  {journal}
  {Proceedings of the National Academy of Sciences}\ }\textbf {\bibinfo
  {volume} {113}},\ \bibinfo {pages} {9177--9181} (\bibinfo {year} {2016})},\
  \Eprint
  {http://arxiv.org/abs/https://www.pnas.org/content/113/33/9177.full.pdf}
  {https://www.pnas.org/content/113/33/9177.full.pdf} \BibitemShut {NoStop}%
\bibitem [{\citenamefont {Böhmer}\ and\ \citenamefont
  {Kreisel}(2017)}]{B_hmer_2017}%
  \BibitemOpen
  \bibfield  {author} {\bibinfo {author} {\bibfnamefont {Anna~E}\ \bibnamefont
  {Böhmer}}\ and\ \bibinfo {author} {\bibfnamefont {Andreas}\ \bibnamefont
  {Kreisel}},\ }\bibfield  {title} {\enquote {\bibinfo {title} {Nematicity,
  magnetism and superconductivity in {FeSe}},}\ }\href {\doibase
  10.1088/1361-648x/aa9caa} {\bibfield  {journal} {\bibinfo  {journal} {Journal
  of Physics: Condensed Matter}\ }\textbf {\bibinfo {volume} {30}},\ \bibinfo
  {pages} {023001} (\bibinfo {year} {2017})}\BibitemShut {NoStop}%
\bibitem [{\citenamefont {Hosoi}\ \emph {et~al.}(2016)\citenamefont {Hosoi},
  \citenamefont {Matsuura}, \citenamefont {Ishida}, \citenamefont {Wang},
  \citenamefont {Mizukami}, \citenamefont {Watashige}, \citenamefont
  {Kasahara}, \citenamefont {Matsuda},\ and\ \citenamefont
  {Shibauchi}}]{Hosoi_2016}%
  \BibitemOpen
  \bibfield  {author} {\bibinfo {author} {\bibfnamefont {Suguru}\ \bibnamefont
  {Hosoi}}, \bibinfo {author} {\bibfnamefont {Kohei}\ \bibnamefont {Matsuura}},
  \bibinfo {author} {\bibfnamefont {Kousuke}\ \bibnamefont {Ishida}}, \bibinfo
  {author} {\bibfnamefont {Hao}\ \bibnamefont {Wang}}, \bibinfo {author}
  {\bibfnamefont {Yuta}\ \bibnamefont {Mizukami}}, \bibinfo {author}
  {\bibfnamefont {Tatsuya}\ \bibnamefont {Watashige}}, \bibinfo {author}
  {\bibfnamefont {Shigeru}\ \bibnamefont {Kasahara}}, \bibinfo {author}
  {\bibfnamefont {Yuji}\ \bibnamefont {Matsuda}}, \ and\ \bibinfo {author}
  {\bibfnamefont {Takasada}\ \bibnamefont {Shibauchi}},\ }\bibfield  {title}
  {\enquote {\bibinfo {title} {Nematic quantum critical point without magnetism
  in {FeSe$_{1-x}$S$_x$} superconductors},}\ }\href {\doibase
  10.1073/pnas.1605806113} {\bibfield  {journal} {\bibinfo  {journal}
  {Proceedings of the National Academy of Sciences}\ }\textbf {\bibinfo
  {volume} {113}},\ \bibinfo {pages} {8139--8143} (\bibinfo {year}
  {2016})}\BibitemShut {NoStop}%
\bibitem [{\citenamefont {Bartlett}\ \emph {et~al.}(2021)\citenamefont
  {Bartlett}, \citenamefont {Steppke}, \citenamefont {Hosoi}, \citenamefont
  {Noad}, \citenamefont {Park}, \citenamefont {Timm}, \citenamefont
  {Shibauchi}, \citenamefont {Mackenzie},\ and\ \citenamefont
  {Hicks}}]{bartlett2021relationship}%
  \BibitemOpen
  \bibfield  {author} {\bibinfo {author} {\bibfnamefont {Jack}\ \bibnamefont
  {Bartlett}}, \bibinfo {author} {\bibfnamefont {Alexander}\ \bibnamefont
  {Steppke}}, \bibinfo {author} {\bibfnamefont {Suguru}\ \bibnamefont {Hosoi}},
  \bibinfo {author} {\bibfnamefont {Hilary}\ \bibnamefont {Noad}}, \bibinfo
  {author} {\bibfnamefont {Joonbum}\ \bibnamefont {Park}}, \bibinfo {author}
  {\bibfnamefont {Carsten}\ \bibnamefont {Timm}}, \bibinfo {author}
  {\bibfnamefont {Takasada}\ \bibnamefont {Shibauchi}}, \bibinfo {author}
  {\bibfnamefont {Andrew~P.}\ \bibnamefont {Mackenzie}}, \ and\ \bibinfo
  {author} {\bibfnamefont {Clifford~W.}\ \bibnamefont {Hicks}},\ }\href@noop {}
  {\enquote {\bibinfo {title} {The relationship between transport anisotropy
  and nematicity in {FeSe}},}\ } (\bibinfo {year} {2021}),\ \Eprint
  {http://arxiv.org/abs/2102.09212} {arXiv:2102.09212 [cond-mat.str-el]}
  \BibitemShut {NoStop}%
\bibitem [{\citenamefont {Wiecki}\ \emph {et~al.}(2017)\citenamefont {Wiecki},
  \citenamefont {Nandi}, \citenamefont {B\"ohmer}, \citenamefont {Bud'ko},
  \citenamefont {Canfield},\ and\ \citenamefont {Furukawa}}]{Wiecki2017}%
  \BibitemOpen
  \bibfield  {author} {\bibinfo {author} {\bibfnamefont {P.}~\bibnamefont
  {Wiecki}}, \bibinfo {author} {\bibfnamefont {M.}~\bibnamefont {Nandi}},
  \bibinfo {author} {\bibfnamefont {A.~E.}\ \bibnamefont {B\"ohmer}}, \bibinfo
  {author} {\bibfnamefont {S.~L.}\ \bibnamefont {Bud'ko}}, \bibinfo {author}
  {\bibfnamefont {P.~C.}\ \bibnamefont {Canfield}}, \ and\ \bibinfo {author}
  {\bibfnamefont {Y.}~\bibnamefont {Furukawa}},\ }\bibfield  {title} {\enquote
  {\bibinfo {title} {{NMR} evidence for static local nematicity and its
  cooperative interplay with low-energy magnetic fluctuations in {FeSe} under
  pressure},}\ }\href {\doibase 10.1103/PhysRevB.96.180502} {\bibfield
  {journal} {\bibinfo  {journal} {Phys. Rev. B}\ }\textbf {\bibinfo {volume}
  {96}},\ \bibinfo {pages} {180502} (\bibinfo {year} {2017})}\BibitemShut
  {NoStop}%
\bibitem [{\citenamefont {Wang}\ \emph {et~al.}(2017)\citenamefont {Wang},
  \citenamefont {Zhou}, \citenamefont {Sun}, \citenamefont {Cui}, \citenamefont
  {Li}, \citenamefont {Lei}, \citenamefont {Wang},\ and\ \citenamefont
  {Yu}}]{Wang2017}%
  \BibitemOpen
  \bibfield  {author} {\bibinfo {author} {\bibfnamefont {P.~S.}\ \bibnamefont
  {Wang}}, \bibinfo {author} {\bibfnamefont {P.}~\bibnamefont {Zhou}}, \bibinfo
  {author} {\bibfnamefont {S.~S.}\ \bibnamefont {Sun}}, \bibinfo {author}
  {\bibfnamefont {Y.}~\bibnamefont {Cui}}, \bibinfo {author} {\bibfnamefont
  {T.~R.}\ \bibnamefont {Li}}, \bibinfo {author} {\bibfnamefont {Hechang}\
  \bibnamefont {Lei}}, \bibinfo {author} {\bibfnamefont {Ziqiang}\ \bibnamefont
  {Wang}}, \ and\ \bibinfo {author} {\bibfnamefont {Weiqiang}\ \bibnamefont
  {Yu}},\ }\bibfield  {title} {\enquote {\bibinfo {title} {Robust
  short-range-ordered nematicity in {FeSe} evidenced by high-pressure {NMR}},}\
  }\href {\doibase 10.1103/PhysRevB.96.094528} {\bibfield  {journal} {\bibinfo
  {journal} {Phys. Rev. B}\ }\textbf {\bibinfo {volume} {96}},\ \bibinfo
  {pages} {094528} (\bibinfo {year} {2017})}\BibitemShut {NoStop}%
\bibitem [{\citenamefont {Li}\ \emph {et~al.}(2020)\citenamefont {Li},
  \citenamefont {Lei}, \citenamefont {Zhao}, \citenamefont {Nie}, \citenamefont
  {Song}, \citenamefont {Zheng}, \citenamefont {Li}, \citenamefont {Kang},
  \citenamefont {Luo}, \citenamefont {Wu},\ and\ \citenamefont
  {Chen}}]{Li2020}%
  \BibitemOpen
  \bibfield  {author} {\bibinfo {author} {\bibfnamefont {J.}~\bibnamefont
  {Li}}, \bibinfo {author} {\bibfnamefont {B.}~\bibnamefont {Lei}}, \bibinfo
  {author} {\bibfnamefont {D.}~\bibnamefont {Zhao}}, \bibinfo {author}
  {\bibfnamefont {L.~P.}\ \bibnamefont {Nie}}, \bibinfo {author} {\bibfnamefont
  {D.~W.}\ \bibnamefont {Song}}, \bibinfo {author} {\bibfnamefont {L.~X.}\
  \bibnamefont {Zheng}}, \bibinfo {author} {\bibfnamefont {S.~J.}\ \bibnamefont
  {Li}}, \bibinfo {author} {\bibfnamefont {B.~L.}\ \bibnamefont {Kang}},
  \bibinfo {author} {\bibfnamefont {X.~G.}\ \bibnamefont {Luo}}, \bibinfo
  {author} {\bibfnamefont {T.}~\bibnamefont {Wu}}, \ and\ \bibinfo {author}
  {\bibfnamefont {X.~H.}\ \bibnamefont {Chen}},\ }\bibfield  {title} {\enquote
  {\bibinfo {title} {Spin-orbital-intertwined nematic state in {FeSe}},}\
  }\href {\doibase 10.1103/PhysRevX.10.011034} {\bibfield  {journal} {\bibinfo
  {journal} {Phys. Rev. X}\ }\textbf {\bibinfo {volume} {10}},\ \bibinfo
  {pages} {011034} (\bibinfo {year} {2020})}\BibitemShut {NoStop}%
\bibitem [{\citenamefont {Koch}\ \emph {et~al.}(2019)\citenamefont {Koch},
  \citenamefont {Konstantinova}, \citenamefont {Abeykoon}, \citenamefont
  {Wang}, \citenamefont {Petrovic}, \citenamefont {Zhu}, \citenamefont
  {Bozin},\ and\ \citenamefont {Billinge}}]{Koch2019}%
  \BibitemOpen
  \bibfield  {author} {\bibinfo {author} {\bibfnamefont {R.~J.}\ \bibnamefont
  {Koch}}, \bibinfo {author} {\bibfnamefont {T.}~\bibnamefont {Konstantinova}},
  \bibinfo {author} {\bibfnamefont {M.}~\bibnamefont {Abeykoon}}, \bibinfo
  {author} {\bibfnamefont {A.}~\bibnamefont {Wang}}, \bibinfo {author}
  {\bibfnamefont {C.}~\bibnamefont {Petrovic}}, \bibinfo {author}
  {\bibfnamefont {Y.}~\bibnamefont {Zhu}}, \bibinfo {author} {\bibfnamefont
  {E.~S.}\ \bibnamefont {Bozin}}, \ and\ \bibinfo {author} {\bibfnamefont
  {S.~J.~L.}\ \bibnamefont {Billinge}},\ }\bibfield  {title} {\enquote
  {\bibinfo {title} {Room temperature local nematicity in {FeSe}
  superconductor},}\ }\href {\doibase 10.1103/PhysRevB.100.020501} {\bibfield
  {journal} {\bibinfo  {journal} {Phys. Rev. B}\ }\textbf {\bibinfo {volume}
  {100}},\ \bibinfo {pages} {020501} (\bibinfo {year} {2019})}\BibitemShut
  {NoStop}%
\bibitem [{\citenamefont {Frandsen}\ \emph {et~al.}(2019)\citenamefont
  {Frandsen}, \citenamefont {Wang}, \citenamefont {Wu}, \citenamefont {Zhao},\
  and\ \citenamefont {Birgeneau}}]{Frandsen2019}%
  \BibitemOpen
  \bibfield  {author} {\bibinfo {author} {\bibfnamefont {Benjamin~A.}\
  \bibnamefont {Frandsen}}, \bibinfo {author} {\bibfnamefont {Qisi}\
  \bibnamefont {Wang}}, \bibinfo {author} {\bibfnamefont {Shan}\ \bibnamefont
  {Wu}}, \bibinfo {author} {\bibfnamefont {Jun}\ \bibnamefont {Zhao}}, \ and\
  \bibinfo {author} {\bibfnamefont {Robert~J.}\ \bibnamefont {Birgeneau}},\
  }\bibfield  {title} {\enquote {\bibinfo {title} {Quantitative
  characterization of short-range orthorhombic fluctuations in {FeSe} through
  pair distribution function analysis},}\ }\href {\doibase
  10.1103/PhysRevB.100.020504} {\bibfield  {journal} {\bibinfo  {journal}
  {Phys. Rev. B}\ }\textbf {\bibinfo {volume} {100}},\ \bibinfo {pages}
  {020504} (\bibinfo {year} {2019})}\BibitemShut {NoStop}%
\bibitem [{\citenamefont {Wang}\ \emph {et~al.}(2020)\citenamefont {Wang},
  \citenamefont {Zhao}, \citenamefont {Koch}, \citenamefont {Billinge},\ and\
  \citenamefont {Zunger}}]{Wang2020}%
  \BibitemOpen
  \bibfield  {author} {\bibinfo {author} {\bibfnamefont {Zhi}\ \bibnamefont
  {Wang}}, \bibinfo {author} {\bibfnamefont {Xin-Gang}\ \bibnamefont {Zhao}},
  \bibinfo {author} {\bibfnamefont {Robert}\ \bibnamefont {Koch}}, \bibinfo
  {author} {\bibfnamefont {Simon J.~L.}\ \bibnamefont {Billinge}}, \ and\
  \bibinfo {author} {\bibfnamefont {Alex}\ \bibnamefont {Zunger}},\ }\bibfield
  {title} {\enquote {\bibinfo {title} {Understanding electronic peculiarities
  in tetragonal {FeSe} as local structural symmetry breaking},}\ }\href
  {\doibase 10.1103/PhysRevB.102.235121} {\bibfield  {journal} {\bibinfo
  {journal} {Phys. Rev. B}\ }\textbf {\bibinfo {volume} {102}},\ \bibinfo
  {pages} {235121} (\bibinfo {year} {2020})}\BibitemShut {NoStop}%
\bibitem [{\citenamefont {Massat}\ \emph {et~al.}(2018)\citenamefont {Massat},
  \citenamefont {Quan}, \citenamefont {Grasset}, \citenamefont {M\'easson},
  \citenamefont {Cazayous}, \citenamefont {Sacuto}, \citenamefont {Karlsson},
  \citenamefont {Strobel}, \citenamefont {Toulemonde}, \citenamefont {Yin},\
  and\ \citenamefont {Gallais}}]{Massat2018}%
  \BibitemOpen
  \bibfield  {author} {\bibinfo {author} {\bibfnamefont {Pierre}\ \bibnamefont
  {Massat}}, \bibinfo {author} {\bibfnamefont {Yundi}\ \bibnamefont {Quan}},
  \bibinfo {author} {\bibfnamefont {Romain}\ \bibnamefont {Grasset}}, \bibinfo
  {author} {\bibfnamefont {Marie-Aude}\ \bibnamefont {M\'easson}}, \bibinfo
  {author} {\bibfnamefont {Maximilien}\ \bibnamefont {Cazayous}}, \bibinfo
  {author} {\bibfnamefont {Alain}\ \bibnamefont {Sacuto}}, \bibinfo {author}
  {\bibfnamefont {Sandra}\ \bibnamefont {Karlsson}}, \bibinfo {author}
  {\bibfnamefont {Pierre}\ \bibnamefont {Strobel}}, \bibinfo {author}
  {\bibfnamefont {Pierre}\ \bibnamefont {Toulemonde}}, \bibinfo {author}
  {\bibfnamefont {Zhiping}\ \bibnamefont {Yin}}, \ and\ \bibinfo {author}
  {\bibfnamefont {Yann}\ \bibnamefont {Gallais}},\ }\bibfield  {title}
  {\enquote {\bibinfo {title} {Collapse of critical nematic fluctuations in
  {FeSe} under pressure},}\ }\href {\doibase 10.1103/PhysRevLett.121.077001}
  {\bibfield  {journal} {\bibinfo  {journal} {Phys. Rev. Lett.}\ }\textbf
  {\bibinfo {volume} {121}},\ \bibinfo {pages} {077001} (\bibinfo {year}
  {2018})}\BibitemShut {NoStop}%
\bibitem [{\citenamefont {Xu}\ \emph {et~al.}(2008)\citenamefont {Xu},
  \citenamefont {M\"uller},\ and\ \citenamefont {Sachdev}}]{Xu2008}%
  \BibitemOpen
  \bibfield  {author} {\bibinfo {author} {\bibfnamefont {Cenke}\ \bibnamefont
  {Xu}}, \bibinfo {author} {\bibfnamefont {Markus}\ \bibnamefont {M\"uller}}, \
  and\ \bibinfo {author} {\bibfnamefont {Subir}\ \bibnamefont {Sachdev}},\
  }\bibfield  {title} {\enquote {\bibinfo {title} {Ising and spin orders in the
  iron-based superconductors},}\ }\href {\doibase 10.1103/PhysRevB.78.020501}
  {\bibfield  {journal} {\bibinfo  {journal} {Phys. Rev. B}\ }\textbf {\bibinfo
  {volume} {78}},\ \bibinfo {pages} {020501} (\bibinfo {year}
  {2008})}\BibitemShut {NoStop}%
\bibitem [{\citenamefont {Fang}\ \emph {et~al.}(2008)\citenamefont {Fang},
  \citenamefont {Yao}, \citenamefont {Tsai}, \citenamefont {Hu},\ and\
  \citenamefont {Kivelson}}]{Fang2008}%
  \BibitemOpen
  \bibfield  {author} {\bibinfo {author} {\bibfnamefont {Chen}\ \bibnamefont
  {Fang}}, \bibinfo {author} {\bibfnamefont {Hong}\ \bibnamefont {Yao}},
  \bibinfo {author} {\bibfnamefont {Wei-Feng}\ \bibnamefont {Tsai}}, \bibinfo
  {author} {\bibfnamefont {JiangPing}\ \bibnamefont {Hu}}, \ and\ \bibinfo
  {author} {\bibfnamefont {Steven~A.}\ \bibnamefont {Kivelson}},\ }\bibfield
  {title} {\enquote {\bibinfo {title} {Theory of electron nematic order in
  {LaFeAsO}},}\ }\href {\doibase 10.1103/PhysRevB.77.224509} {\bibfield
  {journal} {\bibinfo  {journal} {Phys. Rev. B}\ }\textbf {\bibinfo {volume}
  {77}},\ \bibinfo {pages} {224509} (\bibinfo {year} {2008})}\BibitemShut
  {NoStop}%
\bibitem [{\citenamefont {Fernandes}\ \emph {et~al.}(2010)\citenamefont
  {Fernandes}, \citenamefont {VanBebber}, \citenamefont {Bhattacharya},
  \citenamefont {Chandra}, \citenamefont {Keppens}, \citenamefont {Mandrus},
  \citenamefont {McGuire}, \citenamefont {Sales}, \citenamefont {Sefat},\ and\
  \citenamefont {Schmalian}}]{Fernandes2010}%
  \BibitemOpen
  \bibfield  {author} {\bibinfo {author} {\bibfnamefont {R.~M.}\ \bibnamefont
  {Fernandes}}, \bibinfo {author} {\bibfnamefont {L.~H.}\ \bibnamefont
  {VanBebber}}, \bibinfo {author} {\bibfnamefont {S.}~\bibnamefont
  {Bhattacharya}}, \bibinfo {author} {\bibfnamefont {P.}~\bibnamefont
  {Chandra}}, \bibinfo {author} {\bibfnamefont {V.}~\bibnamefont {Keppens}},
  \bibinfo {author} {\bibfnamefont {D.}~\bibnamefont {Mandrus}}, \bibinfo
  {author} {\bibfnamefont {M.~A.}\ \bibnamefont {McGuire}}, \bibinfo {author}
  {\bibfnamefont {B.~C.}\ \bibnamefont {Sales}}, \bibinfo {author}
  {\bibfnamefont {A.~S.}\ \bibnamefont {Sefat}}, \ and\ \bibinfo {author}
  {\bibfnamefont {J.}~\bibnamefont {Schmalian}},\ }\bibfield  {title} {\enquote
  {\bibinfo {title} {Effects of nematic fluctuations on the elastic properties
  of iron arsenide superconductors},}\ }\href {\doibase
  10.1103/PhysRevLett.105.157003} {\bibfield  {journal} {\bibinfo  {journal}
  {Phys. Rev. Lett.}\ }\textbf {\bibinfo {volume} {105}},\ \bibinfo {pages}
  {157003} (\bibinfo {year} {2010})}\BibitemShut {NoStop}%
\bibitem [{\citenamefont {Fernandes}\ \emph {et~al.}(2014)\citenamefont
  {Fernandes}, \citenamefont {Chubukov},\ and\ \citenamefont
  {Schmalian}}]{Fernandes2014}%
  \BibitemOpen
  \bibfield  {author} {\bibinfo {author} {\bibfnamefont {R.~M.}\ \bibnamefont
  {Fernandes}}, \bibinfo {author} {\bibfnamefont {A.~V.}\ \bibnamefont
  {Chubukov}}, \ and\ \bibinfo {author} {\bibfnamefont {J.}~\bibnamefont
  {Schmalian}},\ }\bibfield  {title} {\enquote {\bibinfo {title} {What drives
  nematic order in iron-based~superconductors?}}\ }\href {\doibase
  10.1038/nphys2877} {\bibfield  {journal} {\bibinfo  {journal} {Nature
  Physics}\ }\textbf {\bibinfo {volume} {10}},\ \bibinfo {pages} {97--104}
  (\bibinfo {year} {2014})}\BibitemShut {NoStop}%
\bibitem [{\citenamefont {Kr\"uger}\ \emph {et~al.}(2009)\citenamefont
  {Kr\"uger}, \citenamefont {Kumar}, \citenamefont {Zaanen},\ and\
  \citenamefont {van~den Brink}}]{Krueger2009}%
  \BibitemOpen
  \bibfield  {author} {\bibinfo {author} {\bibfnamefont {Frank}\ \bibnamefont
  {Kr\"uger}}, \bibinfo {author} {\bibfnamefont {Sanjeev}\ \bibnamefont
  {Kumar}}, \bibinfo {author} {\bibfnamefont {Jan}\ \bibnamefont {Zaanen}}, \
  and\ \bibinfo {author} {\bibfnamefont {Jeroen}\ \bibnamefont {van~den
  Brink}},\ }\bibfield  {title} {\enquote {\bibinfo {title} {Spin-orbital
  frustrations and anomalous metallic state in iron-pnictide
  superconductors},}\ }\href {\doibase 10.1103/PhysRevB.79.054504} {\bibfield
  {journal} {\bibinfo  {journal} {Phys. Rev. B}\ }\textbf {\bibinfo {volume}
  {79}},\ \bibinfo {pages} {054504} (\bibinfo {year} {2009})}\BibitemShut
  {NoStop}%
\bibitem [{\citenamefont {Lv}\ \emph {et~al.}(2009)\citenamefont {Lv},
  \citenamefont {Wu},\ and\ \citenamefont {Phillips}}]{Lv2009}%
  \BibitemOpen
  \bibfield  {author} {\bibinfo {author} {\bibfnamefont {Weicheng}\
  \bibnamefont {Lv}}, \bibinfo {author} {\bibfnamefont {Jiansheng}\
  \bibnamefont {Wu}}, \ and\ \bibinfo {author} {\bibfnamefont {Philip}\
  \bibnamefont {Phillips}},\ }\bibfield  {title} {\enquote {\bibinfo {title}
  {Orbital ordering induces structural phase transition and the resistivity
  anomaly in iron pnictides},}\ }\href {\doibase 10.1103/PhysRevB.80.224506}
  {\bibfield  {journal} {\bibinfo  {journal} {Phys. Rev. B}\ }\textbf {\bibinfo
  {volume} {80}},\ \bibinfo {pages} {224506} (\bibinfo {year}
  {2009})}\BibitemShut {NoStop}%
\bibitem [{\citenamefont {Chen}\ \emph {et~al.}(2010)\citenamefont {Chen},
  \citenamefont {Maciejko}, \citenamefont {Sorini}, \citenamefont {Moritz},
  \citenamefont {Singh},\ and\ \citenamefont {Devereaux}}]{Chen2010}%
  \BibitemOpen
  \bibfield  {author} {\bibinfo {author} {\bibfnamefont {C.-C.}\ \bibnamefont
  {Chen}}, \bibinfo {author} {\bibfnamefont {J.}~\bibnamefont {Maciejko}},
  \bibinfo {author} {\bibfnamefont {A.~P.}\ \bibnamefont {Sorini}}, \bibinfo
  {author} {\bibfnamefont {B.}~\bibnamefont {Moritz}}, \bibinfo {author}
  {\bibfnamefont {R.~R.~P.}\ \bibnamefont {Singh}}, \ and\ \bibinfo {author}
  {\bibfnamefont {T.~P.}\ \bibnamefont {Devereaux}},\ }\bibfield  {title}
  {\enquote {\bibinfo {title} {Orbital order and spontaneous orthorhombicity in
  iron pnictides},}\ }\href {\doibase 10.1103/PhysRevB.82.100504} {\bibfield
  {journal} {\bibinfo  {journal} {Phys. Rev. B}\ }\textbf {\bibinfo {volume}
  {82}},\ \bibinfo {pages} {100504} (\bibinfo {year} {2010})}\BibitemShut
  {NoStop}%
\bibitem [{\citenamefont {Kontani}\ and\ \citenamefont
  {Onari}(2010)}]{Kontani2010}%
  \BibitemOpen
  \bibfield  {author} {\bibinfo {author} {\bibfnamefont {Hiroshi}\ \bibnamefont
  {Kontani}}\ and\ \bibinfo {author} {\bibfnamefont {Seiichiro}\ \bibnamefont
  {Onari}},\ }\bibfield  {title} {\enquote {\bibinfo {title}
  {Orbital-fluctuation-mediated superconductivity in iron pnictides: Analysis
  of the five-orbital {Hubbard-Holstein} model},}\ }\href {\doibase
  10.1103/PhysRevLett.104.157001} {\bibfield  {journal} {\bibinfo  {journal}
  {Phys. Rev. Lett.}\ }\textbf {\bibinfo {volume} {104}},\ \bibinfo {pages}
  {157001} (\bibinfo {year} {2010})}\BibitemShut {NoStop}%
\bibitem [{\citenamefont {Oganesyan}\ \emph {et~al.}(2001)\citenamefont
  {Oganesyan}, \citenamefont {Kivelson},\ and\ \citenamefont
  {Fradkin}}]{Oganesyan2001}%
  \BibitemOpen
  \bibfield  {author} {\bibinfo {author} {\bibfnamefont {Vadim}\ \bibnamefont
  {Oganesyan}}, \bibinfo {author} {\bibfnamefont {Steven~A.}\ \bibnamefont
  {Kivelson}}, \ and\ \bibinfo {author} {\bibfnamefont {Eduardo}\ \bibnamefont
  {Fradkin}},\ }\bibfield  {title} {\enquote {\bibinfo {title} {Quantum theory
  of a nematic fermi fluid},}\ }\href {\doibase 10.1103/PhysRevB.64.195109}
  {\bibfield  {journal} {\bibinfo  {journal} {Phys. Rev. B}\ }\textbf {\bibinfo
  {volume} {64}},\ \bibinfo {pages} {195109} (\bibinfo {year}
  {2001})}\BibitemShut {NoStop}%
\bibitem [{\citenamefont {Yamase}\ \emph {et~al.}(2005)\citenamefont {Yamase},
  \citenamefont {Oganesyan},\ and\ \citenamefont {Metzner}}]{Yamase2005}%
  \BibitemOpen
  \bibfield  {author} {\bibinfo {author} {\bibfnamefont {Hiroyuki}\
  \bibnamefont {Yamase}}, \bibinfo {author} {\bibfnamefont {Vadim}\
  \bibnamefont {Oganesyan}}, \ and\ \bibinfo {author} {\bibfnamefont {Walter}\
  \bibnamefont {Metzner}},\ }\bibfield  {title} {\enquote {\bibinfo {title}
  {Mean-field theory for symmetry-breaking {Fermi} surface deformations on a
  square lattice},}\ }\href {\doibase 10.1103/PhysRevB.72.035114} {\bibfield
  {journal} {\bibinfo  {journal} {Phys. Rev. B}\ }\textbf {\bibinfo {volume}
  {72}},\ \bibinfo {pages} {035114} (\bibinfo {year} {2005})}\BibitemShut
  {NoStop}%
\bibitem [{\citenamefont {Chubukov}\ \emph {et~al.}(2016)\citenamefont
  {Chubukov}, \citenamefont {Khodas},\ and\ \citenamefont
  {Fernandes}}]{Chubukov2016}%
  \BibitemOpen
  \bibfield  {author} {\bibinfo {author} {\bibfnamefont {Andrey~V.}\
  \bibnamefont {Chubukov}}, \bibinfo {author} {\bibfnamefont {M.}~\bibnamefont
  {Khodas}}, \ and\ \bibinfo {author} {\bibfnamefont {Rafael~M.}\ \bibnamefont
  {Fernandes}},\ }\bibfield  {title} {\enquote {\bibinfo {title} {Magnetism,
  superconductivity, and spontaneous orbital order in iron-based
  superconductors: Which comes first and why?}}\ }\href {\doibase
  10.1103/PhysRevX.6.041045} {\bibfield  {journal} {\bibinfo  {journal} {Phys.
  Rev. X}\ }\textbf {\bibinfo {volume} {6}},\ \bibinfo {pages} {041045}
  (\bibinfo {year} {2016})}\BibitemShut {NoStop}%
\bibitem [{\citenamefont {Nattermann}(1988)}]{Nattermann1988}%
  \BibitemOpen
  \bibfield  {author} {\bibinfo {author} {\bibfnamefont {T.}~\bibnamefont
  {Nattermann}},\ }\bibfield  {title} {\enquote {\bibinfo {title} {Dipolar
  interaction in random-field systems},}\ }\href {\doibase
  https://doi.org/10.1088/0305-4470/21/12/005} {\bibfield  {journal} {\bibinfo
  {journal} {Journal of Phys. A: Math. and General}\ }\textbf {\bibinfo
  {volume} {21}},\ \bibinfo {pages} {6288} (\bibinfo {year}
  {1988})}\BibitemShut {NoStop}%
\bibitem [{\citenamefont {Toh}(1992)}]{Toh1992}%
  \BibitemOpen
  \bibfield  {author} {\bibinfo {author} {\bibfnamefont {H.~S.}\ \bibnamefont
  {Toh}},\ }\bibfield  {title} {\enquote {\bibinfo {title} {Structural phase
  transitions with random strains},}\ }\href {\doibase
  https://doi.org/10.1088/0305-4470/25/18/012} {\bibfield  {journal} {\bibinfo
  {journal} {Journal of Phys. A: Math. and General}\ }\textbf {\bibinfo
  {volume} {25}},\ \bibinfo {pages} {4767} (\bibinfo {year}
  {1992})}\BibitemShut {NoStop}%
\bibitem [{\citenamefont {Fischer}\ and\ \citenamefont
  {Hertz}(1991)}]{spinglass}%
  \BibitemOpen
  \bibfield  {author} {\bibinfo {author} {\bibfnamefont {K.~H.}\ \bibnamefont
  {Fischer}}\ and\ \bibinfo {author} {\bibfnamefont {J.~A.}\ \bibnamefont
  {Hertz}},\ }\href@noop {} {\emph {\bibinfo {title} {Spin Glasses, Cambridge
  Studies in Magnetism}}}\ (\bibinfo  {publisher} {Cambridge University
  Press},\ \bibinfo {year} {1991})\BibitemShut {NoStop}%
\bibitem [{\citenamefont {Blinc}(1990)}]{Blinc1990}%
  \BibitemOpen
  \bibfield  {author} {\bibinfo {author} {\bibfnamefont {R.}~\bibnamefont
  {Blinc}},\ }\bibfield  {title} {\enquote {\bibinfo {title} {{NQR} in dipolar
  and quadrupolar glasses},}\ }\href {\doibase 10.1515/zna-1990-3-417}
  {\bibfield  {journal} {\bibinfo  {journal} {Zeitschrift für Naturforschung
  A}\ }\textbf {\bibinfo {volume} {45}},\ \bibinfo {pages} {313--322} (\bibinfo
  {year} {1990})}\BibitemShut {NoStop}%
\bibitem [{\citenamefont {Chen}\ and\ \citenamefont {Ailion}(1990)}]{Chen1991}%
  \BibitemOpen
  \bibfield  {author} {\bibinfo {author} {\bibfnamefont {Songhua}\ \bibnamefont
  {Chen}}\ and\ \bibinfo {author} {\bibfnamefont {David~C.}\ \bibnamefont
  {Ailion}},\ }\bibfield  {title} {\enquote {\bibinfo {title} {{NMR
  determination of the Edwards-Anderson order parameter in the deuterated
  pseudo-spin-glass
  ${\mathrm{Rb}}_{1\mathrm{\ensuremath{-}}\mathit{x}}$(${\mathrm{ND}}_{4}{)}_{\mathit{x}}{\mathrm{D}}_{2}{\mathrm{PO}}_{4}$:
  Anisotropy and concentration dependence of the ${\mathrm{ND}}_{4}^{+}$
  deuteron second moment}},}\ }\href {\doibase 10.1103/PhysRevB.42.5945}
  {\bibfield  {journal} {\bibinfo  {journal} {Phys. Rev. B}\ }\textbf {\bibinfo
  {volume} {42}},\ \bibinfo {pages} {5945--5952} (\bibinfo {year}
  {1990})}\BibitemShut {NoStop}%
\bibitem [{\citenamefont {Papantopoulos}\ \emph {et~al.}(1994)\citenamefont
  {Papantopoulos}, \citenamefont {Papavassiliou}, \citenamefont {Milia},
  \citenamefont {Schmidt}, \citenamefont {Drumheller}, \citenamefont {Pinto},
  \citenamefont {Blinc},\ and\ \citenamefont {Zalar}}]{Papantopoulos1994}%
  \BibitemOpen
  \bibfield  {author} {\bibinfo {author} {\bibfnamefont {G.}~\bibnamefont
  {Papantopoulos}}, \bibinfo {author} {\bibfnamefont {G.}~\bibnamefont
  {Papavassiliou}}, \bibinfo {author} {\bibfnamefont {F.}~\bibnamefont
  {Milia}}, \bibinfo {author} {\bibfnamefont {V.~H.}\ \bibnamefont {Schmidt}},
  \bibinfo {author} {\bibfnamefont {J.~E.}\ \bibnamefont {Drumheller}},
  \bibinfo {author} {\bibfnamefont {N.~J.}\ \bibnamefont {Pinto}}, \bibinfo
  {author} {\bibfnamefont {R.}~\bibnamefont {Blinc}}, \ and\ \bibinfo {author}
  {\bibfnamefont {B.}~\bibnamefont {Zalar}},\ }\bibfield  {title} {\enquote
  {\bibinfo {title} {$^{75}\mathrm{As}$ nuclear quadrupole resonance in weakly
  substitutionally disordered
  {${\mathrm{Rb}}_{1\ensuremath{-}x}{({\mathrm{NH}}_{4})}_{x}{\mathrm{H}}_{2}\mathrm{As}{\mathrm{O}}_{4}$}},}\
  }\href {\doibase 10.1103/PhysRevLett.73.276} {\bibfield  {journal} {\bibinfo
  {journal} {Phys. Rev. Lett.}\ }\textbf {\bibinfo {volume} {73}},\ \bibinfo
  {pages} {276--279} (\bibinfo {year} {1994})}\BibitemShut {NoStop}%
\bibitem [{\citenamefont {Blinc}\ \emph {et~al.}(1999)\citenamefont {Blinc},
  \citenamefont {Dolin\ifmmode~\check{s}\else \v{s}\fi{}ek}, \citenamefont
  {Gregorovi\ifmmode~\check{c}\else \v{c}\fi{}}, \citenamefont {Zalar},
  \citenamefont {Filipi\ifmmode~\check{c}\else \v{c}\fi{}}, \citenamefont
  {Kutnjak}, \citenamefont {Levstik},\ and\ \citenamefont {Pirc}}]{Blinc1999}%
  \BibitemOpen
  \bibfield  {author} {\bibinfo {author} {\bibfnamefont {R.}~\bibnamefont
  {Blinc}}, \bibinfo {author} {\bibfnamefont {J.}~\bibnamefont
  {Dolin\ifmmode~\check{s}\else \v{s}\fi{}ek}}, \bibinfo {author}
  {\bibfnamefont {A.}~\bibnamefont {Gregorovi\ifmmode~\check{c}\else
  \v{c}\fi{}}}, \bibinfo {author} {\bibfnamefont {B.}~\bibnamefont {Zalar}},
  \bibinfo {author} {\bibfnamefont {C.}~\bibnamefont
  {Filipi\ifmmode~\check{c}\else \v{c}\fi{}}}, \bibinfo {author} {\bibfnamefont
  {Z.}~\bibnamefont {Kutnjak}}, \bibinfo {author} {\bibfnamefont
  {A.}~\bibnamefont {Levstik}}, \ and\ \bibinfo {author} {\bibfnamefont
  {R.}~\bibnamefont {Pirc}},\ }\bibfield  {title} {\enquote {\bibinfo {title}
  {Local polarization distribution and {Edwards-Anderson} order parameter of
  relaxor ferroelectrics},}\ }\href {\doibase 10.1103/PhysRevLett.83.424}
  {\bibfield  {journal} {\bibinfo  {journal} {Phys. Rev. Lett.}\ }\textbf
  {\bibinfo {volume} {83}},\ \bibinfo {pages} {424--427} (\bibinfo {year}
  {1999})}\BibitemShut {NoStop}%
\bibitem [{\citenamefont {Karahasanovic}\ and\ \citenamefont
  {Schmalian}(2016)}]{Karahasanovic2016}%
  \BibitemOpen
  \bibfield  {author} {\bibinfo {author} {\bibfnamefont {U.}~\bibnamefont
  {Karahasanovic}}\ and\ \bibinfo {author} {\bibfnamefont {J.}~\bibnamefont
  {Schmalian}},\ }\bibfield  {title} {\enquote {\bibinfo {title} {Elastic
  coupling and spin-driven nematicity in iron-based superconductors},}\ }\href
  {\doibase 10.1103/PhysRevB.93.064520} {\bibfield  {journal} {\bibinfo
  {journal} {Phys. Rev. B}\ }\textbf {\bibinfo {volume} {93}},\ \bibinfo
  {pages} {064520} (\bibinfo {year} {2016})}\BibitemShut {NoStop}%
\bibitem [{\citenamefont {Schneider}\ and\ \citenamefont
  {Pytte}(1977)}]{Schneider1977}%
  \BibitemOpen
  \bibfield  {author} {\bibinfo {author} {\bibfnamefont {T.}~\bibnamefont
  {Schneider}}\ and\ \bibinfo {author} {\bibfnamefont {E.}~\bibnamefont
  {Pytte}},\ }\bibfield  {title} {\enquote {\bibinfo {title} {Random-field
  instability of the ferromagnetic state},}\ }\href {\doibase
  10.1103/PhysRevB.15.1519} {\bibfield  {journal} {\bibinfo  {journal} {Phys.
  Rev. B}\ }\textbf {\bibinfo {volume} {15}},\ \bibinfo {pages} {1519--1522}
  (\bibinfo {year} {1977})}\BibitemShut {NoStop}%
\bibitem [{\citenamefont {Gallais}\ and\ \citenamefont
  {Paul}(2016)}]{Gallais2016}%
  \BibitemOpen
  \bibfield  {author} {\bibinfo {author} {\bibfnamefont {Yann}\ \bibnamefont
  {Gallais}}\ and\ \bibinfo {author} {\bibfnamefont {Indranil}\ \bibnamefont
  {Paul}},\ }\bibfield  {title} {\enquote {\bibinfo {title} {Charge nematicity
  and electronic {Raman} scattering in iron-based superconductors},}\ }\href
  {\doibase 10.1016/j.crhy.2015.10.001} {\bibfield  {journal} {\bibinfo
  {journal} {Comptes Rendus Physique}\ }\textbf {\bibinfo {volume} {17}},\
  \bibinfo {pages} {113--139} (\bibinfo {year} {2016})}\BibitemShut {NoStop}%
\bibitem [{\citenamefont {Kothapalli}\ \emph {et~al.}(2016)\citenamefont
  {Kothapalli}, \citenamefont {Böhmer}, \citenamefont {Jayasekara},
  \citenamefont {Ueland}, \citenamefont {Das}, \citenamefont {Sapkota},
  \citenamefont {Taufour}, \citenamefont {Xiao}, \citenamefont {Alp},
  \citenamefont {Bud'ko}, \citenamefont {Canfield}, \citenamefont {Kreyssig},\
  and\ \citenamefont {Goldman}}]{Kothapalli_2016}%
  \BibitemOpen
  \bibfield  {author} {\bibinfo {author} {\bibfnamefont {K.}~\bibnamefont
  {Kothapalli}}, \bibinfo {author} {\bibfnamefont {A.~E.}\ \bibnamefont
  {Böhmer}}, \bibinfo {author} {\bibfnamefont {W.~T.}\ \bibnamefont
  {Jayasekara}}, \bibinfo {author} {\bibfnamefont {B.~G.}\ \bibnamefont
  {Ueland}}, \bibinfo {author} {\bibfnamefont {P.}~\bibnamefont {Das}},
  \bibinfo {author} {\bibfnamefont {A.}~\bibnamefont {Sapkota}}, \bibinfo
  {author} {\bibfnamefont {V.}~\bibnamefont {Taufour}}, \bibinfo {author}
  {\bibfnamefont {Y.}~\bibnamefont {Xiao}}, \bibinfo {author} {\bibfnamefont
  {E.}~\bibnamefont {Alp}}, \bibinfo {author} {\bibfnamefont {S.~L.}\
  \bibnamefont {Bud'ko}}, \bibinfo {author} {\bibfnamefont {P.~C.}\
  \bibnamefont {Canfield}}, \bibinfo {author} {\bibfnamefont {A.}~\bibnamefont
  {Kreyssig}}, \ and\ \bibinfo {author} {\bibfnamefont {A.~I.}\ \bibnamefont
  {Goldman}},\ }\bibfield  {title} {\enquote {\bibinfo {title} {Strong
  cooperative coupling of pressure-induced magnetic order and nematicity in
  {FeSe}},}\ }\href {\doibase 10.1038/ncomms12728} {\bibfield  {journal}
  {\bibinfo  {journal} {Nature Communications}\ }\textbf {\bibinfo {volume}
  {7}} (\bibinfo {year} {2016}),\ 10.1038/ncomms12728}\BibitemShut {NoStop}%
\bibitem [{\citenamefont {He}\ \emph {et~al.}(2018)\citenamefont {He},
  \citenamefont {Wang}, \citenamefont {Hardy}, \citenamefont {Xu},
  \citenamefont {Wolf}, \citenamefont {Adelmann},\ and\ \citenamefont
  {Meingast}}]{He2018}%
  \BibitemOpen
  \bibfield  {author} {\bibinfo {author} {\bibfnamefont {Mingquan}\
  \bibnamefont {He}}, \bibinfo {author} {\bibfnamefont {Liran}\ \bibnamefont
  {Wang}}, \bibinfo {author} {\bibfnamefont {Fr\'ed\'eric}\ \bibnamefont
  {Hardy}}, \bibinfo {author} {\bibfnamefont {Liping}\ \bibnamefont {Xu}},
  \bibinfo {author} {\bibfnamefont {Thomas}\ \bibnamefont {Wolf}}, \bibinfo
  {author} {\bibfnamefont {Peter}\ \bibnamefont {Adelmann}}, \ and\ \bibinfo
  {author} {\bibfnamefont {Christoph}\ \bibnamefont {Meingast}},\ }\bibfield
  {title} {\enquote {\bibinfo {title} {Evidence for short-range magnetic order
  in the nematic phase of {FeSe} from anisotropic in-plane magnetostriction and
  susceptibility measurements},}\ }\href {\doibase 10.1103/PhysRevB.97.104107}
  {\bibfield  {journal} {\bibinfo  {journal} {Phys. Rev. B}\ }\textbf {\bibinfo
  {volume} {97}},\ \bibinfo {pages} {104107} (\bibinfo {year}
  {2018})}\BibitemShut {NoStop}%
\bibitem [{\citenamefont {Zhou}\ \emph {et~al.}(2020)\citenamefont {Zhou},
  \citenamefont {Scherer}, \citenamefont {Mayaffre}, \citenamefont
  {Toulemonde}, \citenamefont {Ma}, \citenamefont {Li}, \citenamefont
  {Andersen},\ and\ \citenamefont {Julien}}]{Zhou_2020}%
  \BibitemOpen
  \bibfield  {author} {\bibinfo {author} {\bibfnamefont {Rui}\ \bibnamefont
  {Zhou}}, \bibinfo {author} {\bibfnamefont {Daniel~D.}\ \bibnamefont
  {Scherer}}, \bibinfo {author} {\bibfnamefont {Hadrien}\ \bibnamefont
  {Mayaffre}}, \bibinfo {author} {\bibfnamefont {Pierre}\ \bibnamefont
  {Toulemonde}}, \bibinfo {author} {\bibfnamefont {Mingwei}\ \bibnamefont
  {Ma}}, \bibinfo {author} {\bibfnamefont {Yuan}\ \bibnamefont {Li}}, \bibinfo
  {author} {\bibfnamefont {Brian~M.}\ \bibnamefont {Andersen}}, \ and\ \bibinfo
  {author} {\bibfnamefont {Marc-Henri}\ \bibnamefont {Julien}},\ }\bibfield
  {title} {\enquote {\bibinfo {title} {Singular magnetic anisotropy in the
  nematic phase of {FeSe}},}\ }\href {\doibase 10.1038/s41535-020-00295-1}
  {\bibfield  {journal} {\bibinfo  {journal} {npj Quantum Materials}\ }\textbf
  {\bibinfo {volume} {5}} (\bibinfo {year} {2020}),\
  10.1038/s41535-020-00295-1}\BibitemShut {NoStop}%
\bibitem [{\citenamefont {Ning}\ \emph {et~al.}(2010)\citenamefont {Ning},
  \citenamefont {Ahilan}, \citenamefont {Imai}, \citenamefont {Sefat},
  \citenamefont {McGuire}, \citenamefont {Sales}, \citenamefont {Mandrus},
  \citenamefont {Cheng}, \citenamefont {Shen},\ and\ \citenamefont
  {Wen}}]{Ning2010}%
  \BibitemOpen
  \bibfield  {author} {\bibinfo {author} {\bibfnamefont {F.~L.}\ \bibnamefont
  {Ning}}, \bibinfo {author} {\bibfnamefont {K.}~\bibnamefont {Ahilan}},
  \bibinfo {author} {\bibfnamefont {T.}~\bibnamefont {Imai}}, \bibinfo {author}
  {\bibfnamefont {A.~S.}\ \bibnamefont {Sefat}}, \bibinfo {author}
  {\bibfnamefont {M.~A.}\ \bibnamefont {McGuire}}, \bibinfo {author}
  {\bibfnamefont {B.~C.}\ \bibnamefont {Sales}}, \bibinfo {author}
  {\bibfnamefont {D.}~\bibnamefont {Mandrus}}, \bibinfo {author} {\bibfnamefont
  {P.}~\bibnamefont {Cheng}}, \bibinfo {author} {\bibfnamefont
  {B.}~\bibnamefont {Shen}}, \ and\ \bibinfo {author} {\bibfnamefont {H.-H}\
  \bibnamefont {Wen}},\ }\bibfield  {title} {\enquote {\bibinfo {title}
  {Contrasting spin dynamics between underdoped and overdoped
  {$\mathrm{Ba}({\mathrm{Fe}}_{1\ensuremath{-}x}{\mathrm{Co}}_{x}{)}_{2}{\mathrm{As}}_{2}$}},}\
  }\href {\doibase 10.1103/PhysRevLett.104.037001} {\bibfield  {journal}
  {\bibinfo  {journal} {Phys. Rev. Lett.}\ }\textbf {\bibinfo {volume} {104}},\
  \bibinfo {pages} {037001} (\bibinfo {year} {2010})}\BibitemShut {NoStop}%
\bibitem [{\citenamefont {Baek}\ \emph {et~al.}(2014)\citenamefont {Baek},
  \citenamefont {Efremov}, \citenamefont {Ok}, \citenamefont {Kim},
  \citenamefont {van~den Brink},\ and\ \citenamefont {Büchner}}]{Baek_2014}%
  \BibitemOpen
  \bibfield  {author} {\bibinfo {author} {\bibfnamefont {S-H.}\ \bibnamefont
  {Baek}}, \bibinfo {author} {\bibfnamefont {D.~V.}\ \bibnamefont {Efremov}},
  \bibinfo {author} {\bibfnamefont {J.~M.}\ \bibnamefont {Ok}}, \bibinfo
  {author} {\bibfnamefont {J.~S.}\ \bibnamefont {Kim}}, \bibinfo {author}
  {\bibfnamefont {Jeroen}\ \bibnamefont {van~den Brink}}, \ and\ \bibinfo
  {author} {\bibfnamefont {B.}~\bibnamefont {Büchner}},\ }\bibfield  {title}
  {\enquote {\bibinfo {title} {Orbital-driven nematicity in {FeSe}},}\ }\href
  {\doibase 10.1038/nmat4138} {\bibfield  {journal} {\bibinfo  {journal}
  {Nature Materials}\ }\textbf {\bibinfo {volume} {14}},\ \bibinfo {pages}
  {210--214} (\bibinfo {year} {2014})}\BibitemShut {NoStop}%
\bibitem [{\citenamefont {Reiss}\ \emph {et~al.}(2021)\citenamefont {Reiss},
  \citenamefont {Graf}, \citenamefont {Haghighirad}, \citenamefont {Vojta},\
  and\ \citenamefont {Coldea}}]{reiss2021signatures}%
  \BibitemOpen
  \bibfield  {author} {\bibinfo {author} {\bibfnamefont {Pascal}\ \bibnamefont
  {Reiss}}, \bibinfo {author} {\bibfnamefont {David}\ \bibnamefont {Graf}},
  \bibinfo {author} {\bibfnamefont {Amir~A.}\ \bibnamefont {Haghighirad}},
  \bibinfo {author} {\bibfnamefont {Thomas}\ \bibnamefont {Vojta}}, \ and\
  \bibinfo {author} {\bibfnamefont {Amalia~I.}\ \bibnamefont {Coldea}},\
  }\href@noop {} {\enquote {\bibinfo {title} {Signatures of a quantum
  {Griffiths} phase close to an electronic nematic quantum phase transition},}\
  } (\bibinfo {year} {2021}),\ \Eprint {http://arxiv.org/abs/2103.07991}
  {arXiv:2103.07991 [cond-mat.str-el]} \BibitemShut {NoStop}%
\bibitem [{\citenamefont {Millis}\ \emph {et~al.}(2001)\citenamefont {Millis},
  \citenamefont {Morr},\ and\ \citenamefont {Schmalian}}]{Millis2001}%
  \BibitemOpen
  \bibfield  {author} {\bibinfo {author} {\bibfnamefont {A.~J.}\ \bibnamefont
  {Millis}}, \bibinfo {author} {\bibfnamefont {D.~K.}\ \bibnamefont {Morr}}, \
  and\ \bibinfo {author} {\bibfnamefont {J.}~\bibnamefont {Schmalian}},\
  }\bibfield  {title} {\enquote {\bibinfo {title} {Local defect in metallic
  quantum critical systems},}\ }\href {\doibase 10.1103/PhysRevLett.87.167202}
  {\bibfield  {journal} {\bibinfo  {journal} {Phys. Rev. Lett.}\ }\textbf
  {\bibinfo {volume} {87}},\ \bibinfo {pages} {167202} (\bibinfo {year}
  {2001})}\BibitemShut {NoStop}%
\bibitem [{\citenamefont {Millis}\ \emph {et~al.}(2002)\citenamefont {Millis},
  \citenamefont {Morr},\ and\ \citenamefont {Schmalian}}]{Millis2002}%
  \BibitemOpen
  \bibfield  {author} {\bibinfo {author} {\bibfnamefont {A.~J.}\ \bibnamefont
  {Millis}}, \bibinfo {author} {\bibfnamefont {D.~K.}\ \bibnamefont {Morr}}, \
  and\ \bibinfo {author} {\bibfnamefont {J.}~\bibnamefont {Schmalian}},\
  }\bibfield  {title} {\enquote {\bibinfo {title} {Quantum {Griffiths} effects
  in metallic systems},}\ }\href {\doibase 10.1103/PhysRevB.66.174433}
  {\bibfield  {journal} {\bibinfo  {journal} {Phys. Rev. B}\ }\textbf {\bibinfo
  {volume} {66}},\ \bibinfo {pages} {174433} (\bibinfo {year}
  {2002})}\BibitemShut {NoStop}%
\bibitem [{\citenamefont {Vojta}(2003)}]{Vojta2003}%
  \BibitemOpen
  \bibfield  {author} {\bibinfo {author} {\bibfnamefont {Thomas}\ \bibnamefont
  {Vojta}},\ }\bibfield  {title} {\enquote {\bibinfo {title} {Disorder-induced
  rounding of certain quantum phase transitions},}\ }\href {\doibase
  10.1103/PhysRevLett.90.107202} {\bibfield  {journal} {\bibinfo  {journal}
  {Phys. Rev. Lett.}\ }\textbf {\bibinfo {volume} {90}},\ \bibinfo {pages}
  {107202} (\bibinfo {year} {2003})}\BibitemShut {NoStop}%
\bibitem [{\citenamefont {Vojta}\ and\ \citenamefont
  {Schmalian}(2005)}]{Vojta2005}%
  \BibitemOpen
  \bibfield  {author} {\bibinfo {author} {\bibfnamefont {Thomas}\ \bibnamefont
  {Vojta}}\ and\ \bibinfo {author} {\bibfnamefont {J\"org}\ \bibnamefont
  {Schmalian}},\ }\bibfield  {title} {\enquote {\bibinfo {title} {Quantum
  {Griffiths} effects in itinerant {Heisenberg} magnets},}\ }\href {\doibase
  10.1103/PhysRevB.72.045438} {\bibfield  {journal} {\bibinfo  {journal} {Phys.
  Rev. B}\ }\textbf {\bibinfo {volume} {72}},\ \bibinfo {pages} {045438}
  (\bibinfo {year} {2005})}\BibitemShut {NoStop}%
\bibitem [{\citenamefont {Gati}\ \emph {et~al.}(2020)\citenamefont {Gati},
  \citenamefont {Xiang}, \citenamefont {Bud'ko},\ and\ \citenamefont
  {Canfield}}]{Gati_2020}%
  \BibitemOpen
  \bibfield  {author} {\bibinfo {author} {\bibfnamefont {Elena}\ \bibnamefont
  {Gati}}, \bibinfo {author} {\bibfnamefont {Li}~\bibnamefont {Xiang}},
  \bibinfo {author} {\bibfnamefont {Sergey~L.}\ \bibnamefont {Bud'ko}}, \ and\
  \bibinfo {author} {\bibfnamefont {Paul~C.}\ \bibnamefont {Canfield}},\
  }\bibfield  {title} {\enquote {\bibinfo {title} {Measurements of
  elastoresistance under pressure by combining in-situ tunable quasi-uniaxial
  stress with hydrostatic pressure},}\ }\href {\doibase 10.1063/1.5139437}
  {\bibfield  {journal} {\bibinfo  {journal} {Review of Scientific
  Instruments}\ }\textbf {\bibinfo {volume} {91}},\ \bibinfo {pages} {023904}
  (\bibinfo {year} {2020})}\BibitemShut {NoStop}%
\end{thebibliography}%
\end{document}